\documentclass[12pt,prb,onecolumn,superscriptaddress,preprintnumbers,amsmath,amssymb,floatfix]{revtex4-2}

\usepackage{pgf}
\usepackage{graphics,amssymb,amsmath,epsfig,color}

\usepackage{dcolumn}
\usepackage{bm}
\usepackage{float}
\usepackage{natbib}

\usepackage[export]{adjustbox}

\usepackage{multirow}

\usepackage{braket}
\setcitestyle{square}

\usepackage{graphicx}
\usepackage{dcolumn}
\usepackage{bm}
\usepackage{float}
\usepackage{natbib}
\setcitestyle{square}

\usepackage{xcolor}

\usepackage{hyperref}
\hypersetup{colorlinks=true, linkcolor=blue, citecolor=blue, urlcolor=blue}

\newcommand{\beginsupplement}{%
        \setcounter{table}{0}
        \renewcommand{\thetable}{S\arabic{table}}%
        \setcounter{figure}{0}
        \renewcommand{\thefigure}{S\arabic{figure}}%
     }

\usepackage{setspace}

\begin{document}


\title{Large and tunable spin-orbit effect of 6p orbitals through structural cavities in crystals}

\author{Mauro Fava}
\affiliation{Physique Th\'eorique des Mat\'eriaux, QMAT, CESAM, Universit\'e de Li\`ege, B-4000 Sart-Tilman, Belgium}

\author{William Lafargue-Dit-Hauret}
\affiliation{Physique Th\'eorique des Mat\'eriaux, QMAT, CESAM, Universit\'e de Li\`ege, B-4000 Sart-Tilman, Belgium}
\affiliation{Universite de Pau et des Pays de l’Adour, E2S UPPA, CNRS, IPREM, Pau, France}


\author{Aldo H Romero}
\affiliation{Department of Physics and Astronomy, West Virginia University, Morgantown, WV 26505-6315, USA}

\author{Eric Bousquet}
\affiliation{Physique Th\'eorique des Mat\'eriaux, QMAT, CESAM, Universit\'e de Li\`ege, B-4000 Sart-Tilman, Belgium}

\date{\today}

\begin{abstract}
We explore from first-principles calculations the ferroelectric material Pb$_5$Ge$_3$O$_{11}$ as a model for controlling the spin-orbit interaction (SOC) in crystalline solids.
The SOC has a surprisingly strong effect on the structural energy landscape by deepening the ferroelectric double well.
We observe that this effect comes from a specific Pb Wyckoff site that lies on the verge of a natural cavity channel of the crystal.
We also find that a unique cavity state is formed by the empty 6p states of another Pb site at the edge of the cavity channel.
This cavity state exhibits a sizeable spin splitting with a mixed Rashba-Weyl character and a topologically protected crossing of the related bands.
We also show that the ferroelectric properties and the significant SOC effects are exceptionally robust against n-doping up to several electrons per unit cell.
We trace the provenance of these original effects to the unique combination of the structural cavity channel and the chemistry of the Pb atoms with 6p orbitals localizing inside the channel.
\end{abstract}

\maketitle

Relativistic atomic spin-orbit coupling (ASOC) was first introduced in the early 1930s during the development of quantum mechanics. It refers to the interaction between the electronic spin (\textbf{S}) and its angular momentum (\textbf{L}).
Even though ASOC is weak compared with Coulomb or kinetic interactions (one to two orders of magnitude) and even weaker in molecules or crystals owing to the quenching of \textbf{L} with chemical bonding, it appears to be the fundamental interaction to describe, for example, the atomic magnetic moment directions (magnetic anisotropy), magnetostriction or spin canting, and weak ferromagnetism~\cite{stohr2006}.
Hence, SOC has been a centerpiece of molecular and condensed matter physics, and a recent revival of interest is now at play with the discovery of new SOC-related phenomena like spin torques~\cite{RevModPhys.91.035004, torque_2011}, skyrmions \cite{tokura2021, everschor-sitte2018,fert2017}, the presence of a topological Z$_{2}$ order~\cite{topo_review,topo_review_2}, quantum spin-Hall effect~\cite{Murakami2003,Sinova,Kato2004,QSHE_Bernevig}, the existence of spin states with long lifetimes~\cite{Koralek2009,Walser2012,Sasaki2014,Tao2018,Tao_2021}, linear~\cite{Rashba_1960,Yu_Bychkov_1984,di_Sante_Rashba_2013,PhysRevB.95.245141,PhysRev.100.580} and cubic~\cite{PhysRevLett.113.086601,PhysRevLett.108.206601,PhysRevB.94.165202} Rashba (R) and Dresselhaus (D) spin splitting, and so on. 
These new phenomena are significant for future spintronic applications. The term spin-orbitronics was also foreseen when the SOC is the driving ingredient \cite{trier2022}. 

Thus, controlling spin-orbital features is paramount for realizing numerous phenomena with high technological impact.
On the other hand, finding a single material that encompasses several useful and significant SOC features and guarantees, as a matter of principle, a reasonable degree of handling over the "internal" parameters is difficult.
Furthermore, concerning the Rashba physics, which requires doping to be harnessed in polar insulators, a known problem is the preservation of the mirror symmetry breaking in doping conditions since screening by free charges tends to destabilize the electric polarisation~\cite{Djani2019}.
In this letter, we address both issues at once. 
We use the ferroelectric compound lead germanate oxide Pb$_5$Ge$_3$O$_{11}$ (PGO)~\cite{iwasaki1971,Iwasaki1972} as a single platform for the manipulation of spin-orbit interaction. 
We show from density functional theory (DFT) calculations~\cite{gonze2020, Kresse1999} that SOC has an unexpectedly significant impact on both the structural energy landscape of PGO and its electronic structure with a mixed Rashba-Weyl crossing between the spin bands, which is topologically protected by a Z$_2$ invariant. 
More specifically, we show that this significant SOC effect originates from two unique features: i) a vacuum channel in the crystal structure that localizes and unquenches the empty 6p orbitals of some specific lead cations, and ii) the breaking of the mirror site symmetry at other Pb sites. In addition, we show that, unlike common ferroelectric materials, the ferroelectric energy is enhanced by negative carrier doping, which we explain in terms of the short-range nature of the polar instability and localization of the aforementioned 6p states.
From these results, we discuss the design rules for controlling spin-orbital features in solid materials.

\begin{figure}
 \centering
 \includegraphics[width=.75\linewidth]{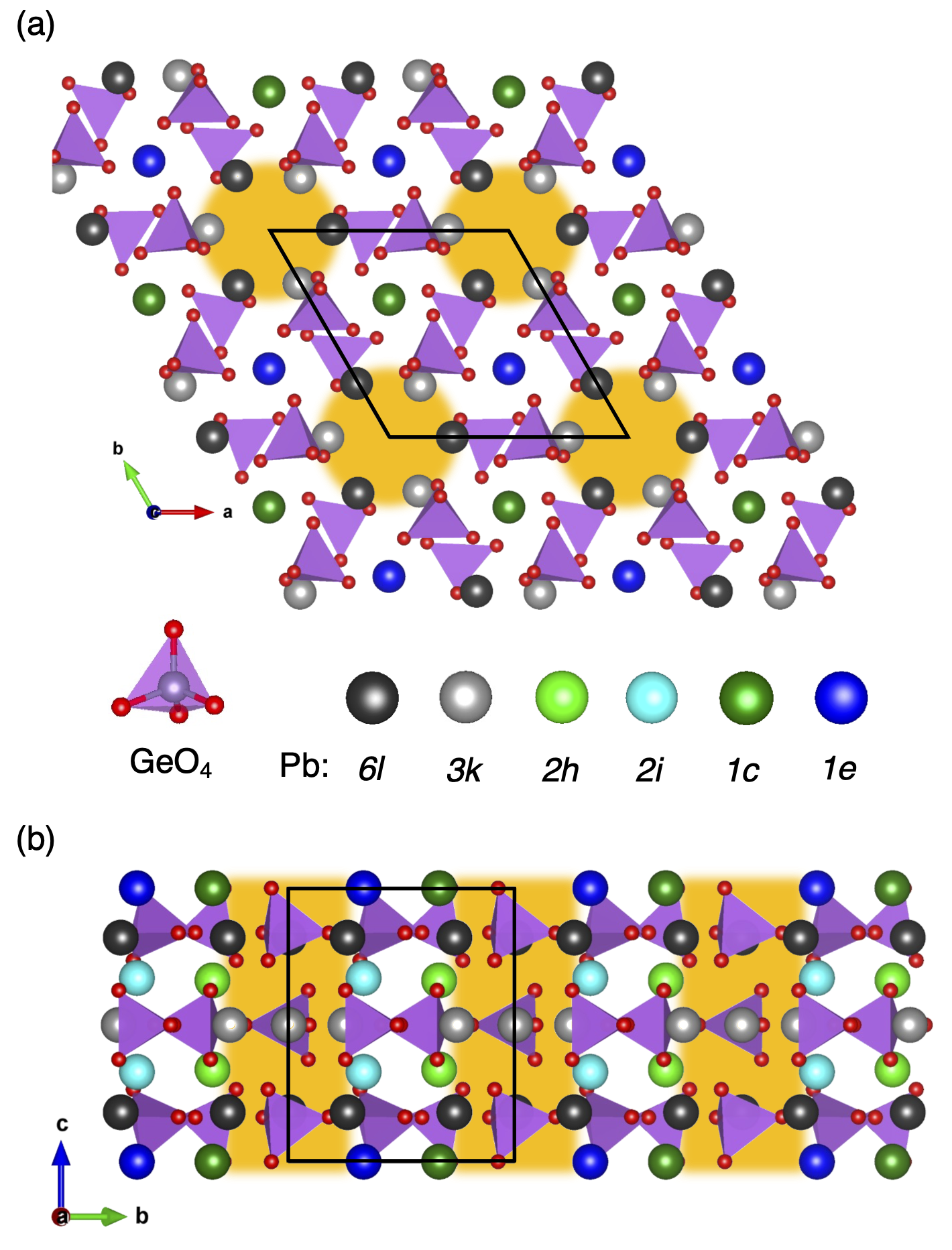}
 \caption{Top (a) and side (b) view of PGO (PE phase). Oxygen atoms are shown in red, Ge atoms and germanate units in purple, while the lead ions are distinguished by their Wyckoff positions. Empty channels are evidenced in gold.}
 \label{fig:PE_struct}
\end{figure}

PGO is a bandgap insulator that undergoes a ferroelectric structural phase transition at 450 K ~\cite{iwasaki1971}. 
Hence, it is a room-temperature ferroelectric (FE) and chiral material ($P3$ space group 143) with a measured spontaneous polarization of $\sim$5 $\mu$C/cm$^{-2}$.  
The combination of chirality and ferroelectricity makes PGO gyroelectric and electrogyroelectric and the natural optical activity can be tuned and switched by an applied electric field following a hysteresis process~\cite{C_Konak_1978,Vlokh_1987}.
We provide a schematic view of the high-symmetry $P\Bar{6}$ phase (space group 174) in \autoref{fig:PE_struct}.
The unit cell of PGO contains 57 atoms, and the PE (FE) phase is described by 15 (23) asymmetric Wyckoff positions (WP).  
The crystal structure can be described as follows. 
The germanium atoms either form - along with the surrounding oxygens - GeO$_{4}$ tetrahedra ($z=0.5$, $6l$ WP) or Ge$_{2}$O$_{7}$ dimers ($z=0$, $3k$ WP)
The lead atoms bridged the Ge$_{2}$O$_{7}$ and GeO$_{4}$ units. 
Pb atoms can be separated into two groups. The first group of Pb atoms was located in $6l$ and $3k$ WP (black and grey atoms in \autoref{fig:PE_struct}) form empty hexagonal channels that propagate along the $[001]$ crystallographic direction (highlighted in yellow in \autoref{fig:PE_struct}). 
The second group consists of Pb atoms found between these channels, that is, the $1e$, $1c$, $2i$, and $2h$ WP (dark blue, dark green, cyan, and lime atoms in \autoref{fig:PE_struct}). Owing to the loss of mirror symmetry in the FE phase, the Pb-$6l$ positions split into two pairs of $3d$ WPs (top and bottom unit cell) in the P3 phase.

Defining $\Delta$E = E(P$\Bar{6}$) - E(P3) as the energy gain between the paraelectric and the ferroelectric phase, we obtain $\Delta$E(no SOC) = 68 meV in the absence of SOC and $\Delta$E(with SOC) = 89 meV when the SOC is included in the calculation, i.e. an increase of 31\%.
This means that the ferroelectric double-well depth of PGO is strongly sensitive to the spin-orbit interaction. 
Furthermore, the SOC enhancement of $\Delta$E is typically not as prominent in lead-based ferroelectrics such as PbTiO$_3$~\cite{arras2019}. 
Thus, these preliminary results call for a deeper investigation of the electronic properties to understand the significant effect of SOC on the ferroelectric well depth. 

\begin{figure*}
\includegraphics[width=17.5cm,keepaspectratio=true]{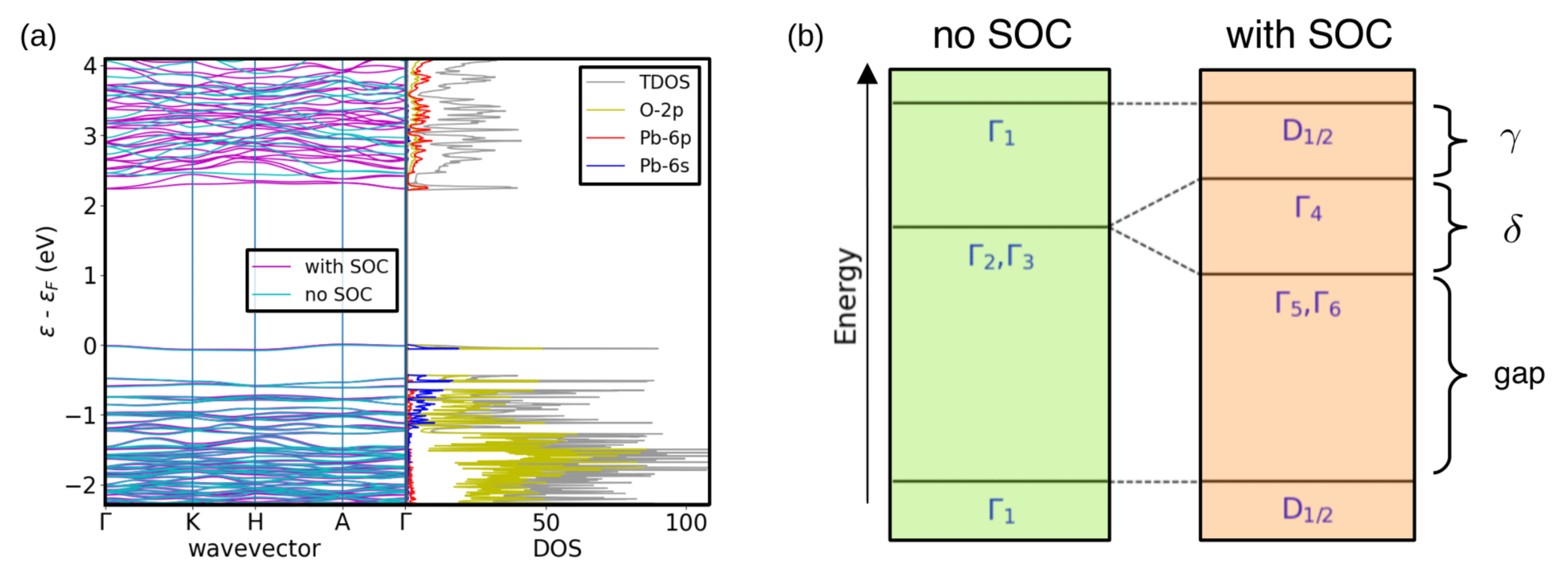}
\caption{(a) Band structure (left) and orbital-projected density of states (right) of the ferroelectric $P3$ phase. The SOC splitting is clearly evident in the conduction bands bottom. (b) A schematics of the spin-orbit induced splitting (single to double irreducible representation) of the CBM / VBM 
levels for the FE phase.
The IR labels of the C$_{3h}$ and C$_{3}$ point groups are the same as in the Bilbao Crystallographic Server, while D is the SO(3) $\times$ $\{1,-1\}$ spin representation reduced to 3-fold rotations and the z-mirror inversion.}
\label{fig:DOS_bands_FE_modified}
 \end{figure*}

In \autoref{fig:DOS_bands_FE_modified} (a), right panel, we report the $spd$ projected DOS around the last occupied valence state and the first unoccupied conduction state of the FE $P3$ phase. 
The top of the valence bands (VB) is dominated by O-2p states followed by contributions from the Pb-6s states and a small amount of Pb-6p states,
suggesting sizable covalent hybridization between the oxygen and lead.
The contributions from the d orbitals are almost absent because, as expected, both Ge and Pb d orbitals are far deeper in energy (approximately -10 eV). 
The conduction bands (CB) are dominated by the Pb-6p spectral weight and show a large Pb-6p/O-2p hybridization (plus the Pb-6s/O-2p in a smaller amount).
In the left panel of \autoref{fig:DOS_bands_FE_modified} (a), we report the electronic band structure of the $P3$ phase in the presence and absence of SOC. 
In the $P3$ ($P\bar{6}$) phase without SOC we obtain a band gap of 2.48 (2.35) eV, which is reduced to 2.25 (2.11) eV if the SOC is included. 
While SOC has only a small effect on the valence band maximum, which has mostly oxygen character, its impact on the CB is sizeable.

To better analyze and quantify the effect of spin-orbit interaction, we perform an irreducible representation (IR) analysis of the VB maximum (VBM) and CB minimum (CBM) states at the $\Gamma$ point. The analysis is reported in section E of the supplementary materials. A scheme of SOC-induced splitting for $P3$ phase is highlighted in \autoref{fig:DOS_bands_FE_modified}(b).
When the SOC is switched off, the top-VB is populated by states belonging to the invariant representation of either the C$_{3h}$ or C$_{3}$ point groups, whereas the bottom-CB is constituted by p$_{x}$, p$_{y}$ orbitals (E' and E single representations of C$_{3h}$ and C$_{3}$ respectively), with a state belonging to the invariant representation IR ($\Gamma_{1}$) located higher in energy. 
With reference to the conduction bands in the P3 phase we define $\gamma = |E(\Bar{\Gamma}_{4}) - E(D_{1/2})|$,
with the split-off energy between the invariant and the p$_x$, p$_y$ orbitals in the absence of SOC as its upper bound.
Clearly, the ferroelectric phase transition does not affect the in-plane p-levels, and adding the spin-orbit results in additional splitting, which in the FE case can be defined as $\delta = |E(\Bar{\Gamma}_{5}\oplus\Bar{\Gamma}_{6}) - E(\Bar{\Gamma}_{4})|$.
From our calculations, we obtain $\delta$ = 180 meV, whereas $\gamma$ is reduced from 270 meV (no SOC) to 106 meV (with SOC).
Such a large SOC effect on the electronic band structure is approximately of the same order of magnitude as that of bulk Au~\cite{rangel2012}, but it is unexpected for ferroelectric insulators with Pb$^{2+}$ cations such as PbTiO$_3$~\cite{arras2019}.



In addition, the orbital angular momentum $\mathbf{L}$ over the $\Gamma$-CBM states is unquenched (as explained in appendix E), which means that the SOC is a first-order correction $\sim \braket{\mathbf{L}}\cdot\mathbf{S}$ of the electronic energies.
We employed the $\mathbf{k}\cdot\mathbf{p}$ approximation near the $\Gamma$ point to further understand the conduction band states. 
The high-symmetry phase has been explored in a previous study~\cite{PhysRevLett.125.216405} therefore, we focus only on the ferroelectric phase. The details of our DFT-based results (both PE and FE cases) are reported in the supplementary materials.

The spin-orbit part of the $P3$ phase ($\Bar{\Gamma}_{5}\oplus\Bar{\Gamma}_{6}$) can be described by:
\begin{equation}\label{FE_k_dot_p}
H^{\text{SOC}}_{\text{P3}}(\mathbf{k}) = \lambda_{R}(k_y\sigma_x - k_x\sigma_y) + \sum_{i=x,y,z}\lambda_{W_i}k_i\sigma_i,
\end{equation}  
where $\lambda_{R}$ (-0.10 eV$\cdot$\AA)
is the Rashba interaction strength and where $\lambda_{W_x}=\lambda_{W_y}\equiv\lambda_{w}$ (0.11 eV$\cdot$\AA) and $\lambda_{W_z}$ (0.01 eV$\cdot$\AA)
represents a Weyl-type band spin splitting, where $\mathbf{\sigma}$ labels the spin. We define $\alpha = \sqrt{\lambda_{R}^2 + \lambda_{w}^2}$ (0.15 eV$\cdot$\AA) to easily quantify the SOC strength.
The values of the SOC parameters is comparable with those of Bi$_{2}$WO$_{6}$, BiAlO$_{3}$, GeTe or BiTeI (Ref.~\cite{Djani2019}), and it is one order of magnitude larger than the values reported in a recent work on LaAlO$_{3}$/LaFeO$_{3}$/SrTiO$_{3}$
~\cite{PhysRevLett.129.187203}. 
Furthermore, exploiting the SOC in this material can be achieved without the need to engineer the unit cell, as it may occur with certain tungsten oxide compounds such as WO$_{3}$, which requires confinement in the direction perpendicular to the polarization ~\cite{Djani2019}. The symmetries of the $P3$ space group allow for an electric field switchable spin texture ~\cite{Tao_2021} and because polar domains of PGO are optically active, this suggests the possibility of controlling the handedness of the spin texture with chiral light. 
It is likely that an electric bias could also be used to tune $\lambda_{R}$, whereas at the same time a magnetic Zeeman interaction may be employed to displace the crossing between the spin bands. In particular, we realized (supplementary section F) that this crossing is protected~\cite{PhysRevB.83.235401} by a Z$_{2}$ topological number because of the presence of a Weyl point at $\Gamma$, which means that the degeneracy of the spin cannot be removed by a magnetic field.

\begin{figure*}
 \centering
 \includegraphics[width=1.\linewidth]{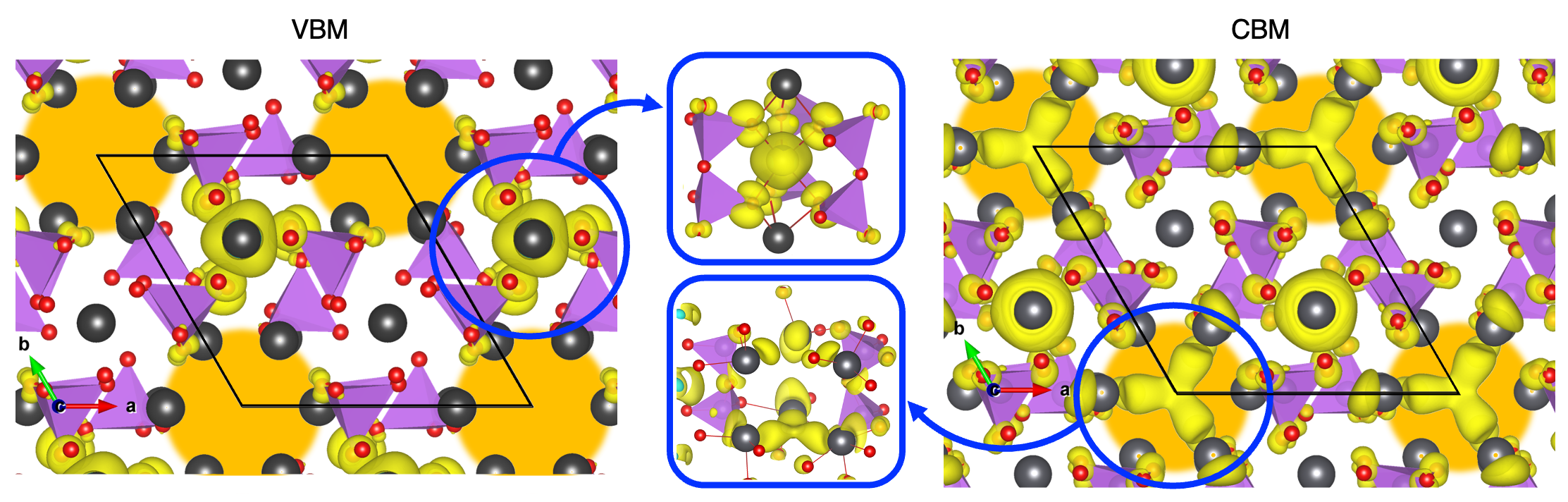}
 \caption{Ferroelectric partial charge density associated with the VBM (iso = 0.001) and CBM (iso = 0.0005) energy windows states (PDOS peaks in \autoref{fig:DOS_bands_FE_modified}). Lead and oxygen atoms are shown in black and red respectively. GeO$_4$ tetrahedra are shown in purple, and empty channels are evidenced in gold.}
 \label{fig:PARCHG_CB_modified}
\end{figure*}

We are left with the need for a microscopic explanation of the large SOC effects. 
We have seen that SOC mainly affects the conduction bands owing to its predominant Pb-6p character. 
To gain further insight, we show in \autoref{fig:PARCHG_CB_modified} the band projected charge density corresponding to the top-VB and bottom-CB isolated bands in the $P3$ phase. 
We found that the top valence electrons were mainly localized at the Pb-1c site and the oxygen 6l sites ($sp$ hybridization), the latter bonding with the Ge$_{2}$O$_{7}$ units and Pb-$1c$ and Pb-$2h$ atoms. This localization near the Ge$_{2}$O$_{7}$ dimers 
is due to the Pb-6s states associated with the steep DOS peak at the Fermi level, which corresponds to $1c$ WP.
On the other hand, the bottom CB charge (where SOC splitting is the most apparent with Pb-6p character) is found to be mostly localized in the vacuum channel and it comes from the Pb-$6l$ WP that are around the cavity.
It is striking to see in \autoref{fig:PARCHG_CB_modified} that this CB of the Pb-$6l$ sites forms a unique and complex cavity state that is quite different from the atomic 6p orbital shapes. It is also interesting to notice that this cavity state exhibits large SOC features, owing to its cavity localization and unquenched $\mathbf{L}$.

To scrutinize the origin of the large SOC observed in PGO, we performed computer experiments by switching the SOC on and off on selected orbitals and at selected atomic sites. 
For each case, we recalculated the ferroelectric double-well depth $\Delta$E and several SOC-related parameters of interest (including the aforementioned $\delta$, $\gamma$ and $\alpha$) and band gap.
The results are reported in tabs. III and IV of the supplementary materials.
This also indicates that the relatively small hybridization of the Pb-6p states with Pb-6s/O-2p is responsible for the SOC renormalization of the energy landscape, as the empty states (CB) do not contribute to the energy.
Now, for the six different Pb WP of the $P\Bar{6}$ phase, we observe that deactivating the SOC at the $3k$ sites significantly affects the PE-FE energy barrier ($\sim$ 17 $\%$ decrease) compared to the $6l$, $1c$, $1e$, $2i$ and $2h$ WPs ($<\sim 7\%$ variation). 
A reason for this - along with the aforementioned Pb-O hybridisation - can be attributed to the fact that the site symmetry group associated with the $3k$ WPs is $m$, which is broken by the phase transition, while the site symmetries induced by the other Pb positions are preserved.
On the other hand, the spin-splitting and split-off CB parameters do not necessarily follow this trend because deactivating the SOC at $1c$, $1e$, $2i$ sites can produce a 2 or 3 times increase in the $\alpha$ parameter (for example, $\Delta$E = 87 meV and $\alpha$ = 2.9 eV$\cdot$\AA in the $2i$-off case). 
Overall, the substitution (alloying) at the selected WP could either affect the ferroelectric domain barrier and/or the CB parameters (spin and band splitting) in a broad fashion, provided that further lowering of the symmetry (if present) and chemical changes have a small impact on the electronic states. 
Hence, PGO with numerous Pb sites and an anisotropic geometry provides an interesting platform for tuning and designing different SOC effects.

\begin{figure}
     \includegraphics[width=0.65\textwidth]{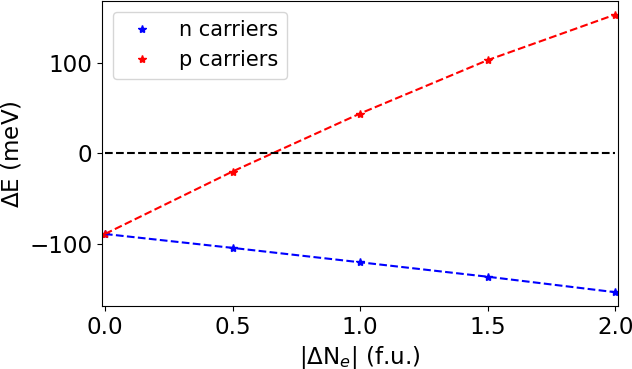}
     \caption{$\Delta$E = E(FE) - E(PE) as a function of extra $n$ and $p$ charge. While an increasing negative carrier concentration lowers the P3 phase energy even further with respect to the paraelectric case, the P$\Bar{6}$ phase can be stabilised instead with a hole concentration above $\sim$ 0.655 f.u..}
     \label{fig:dE_doping}
\end{figure}

Finally, we discuss the robustness of the phase transition under doping. Exploiting the Rashba phenomenology requires, on the one hand, the breaking of the inversion symmetry and, on the other hand, the presence of free carriers, which tend to screen long-range forces responsible for polar instability~\cite{Djani2019,PhysRevLett.109.247601} and thus reduce the magnitude of the spin-splitting parameters. 
We calculated $\Delta$E as a function of $p$ and $n$ carrier doping concentrations for PGO (see \autoref{fig:dE_doping}).
Contrary to regular ferroelectrics~\cite{PhysRevLett.109.247601}, $\Delta$E is surprisingly enhanced by n-doping of the CBM states.
Our calculation of phonons (supplementary section C) reveals that the polar instability has a short-range origin. Hence, the screening of the Coulomb interaction by charges does not affect the instability, as in  BaTiO$_3$~\cite{ghosez1996}.
However, depopulating the VBM (p doping) can stabilize the paraelectric phase above a concentration of $\sim$ 0.66 holes f.u., which highlights the importance of the VBM Pb-6s orbitals for the stabilization of the $P3$ phase.
Finally, we find that the ferromagnetic solution of the n-doping case is lower in energy than the non-magnetic case (supplementary section H), thereby showing that magnetism can occur when doping with electrons. 
When looking at the spin density of these extra electrons, we found that they are indeed strongly localized inside the unique Pb-6p cavity state (see fig.~\ref{fig:spin-density}), making this state, together with the large SOC, a very appealing case study for photoexcitation experiments.


In conclusion, we have shown that the ferroelectric material Pb$_5$Ge$_3$O$_{11}$ can be used as a single platform for controlling diverse spin-orbital properties. 
 We found a large SOC-induced renormalization of ferroelectric double well, which originates from the O-2p/Pb-6p overlap along with the breaking of the mirror site symmetry at the Pb-$3k$ positions. 
Symmetry analysis shows that the FE structure led to mixed Rashba-Weyl spin-splitting.
We argue that the asymmetric localization of the 6p states inside the cavity channel, along with the large Z-number of Pb and first-order nature of the SOC energy correction, can produce large spin-orbital effects.
The deactivation of the SOC at selected WPs also reveals a wide degree of control over the domain barrier and conduction band parameters. 
The localization of the bottom CB levels stems from the presence of natural empty channels and, along with the short-range character of the driving forces of the phase transition, supports ferroelectricity under n-doping conditions. 
Hence, the resulting design rule to obtain large SOC effects in crystals containing Pb$^{2+}$ or Bi$^{3+}$ cations would be to have them placed at the edge of a cavity to form unquenched 6p cavity states. This condition could potentially be explored in other materials with a similar crystal structure where natural empty channels are present, e.g. in Pb$_5$(SiO$_4$)(VO$_4$)$_2$~\cite{krivovichev2004}.
Exploiting the properties of the cavity-confined Pb 6p conduction orbitals would require photo-excitation techniques and/or doping, although alloying as well may be used as an exploratory method. Being relatively confined, these wavefunctions may host novel and unexplored optoelectronic properties.
If we assume a possible dependence on geometrical features, it would be interesting to further explore how the aforementioned states are affected by the size and the shape of the cavity enclosing them.

\section*{Acknowledgements}
The authors aknowledge E. McCabe, Z. Romestan and S. Bandyopadhyay for fruitful discussions. 
Computational resources have been provided by the Consortium des \'Equipements de Calcul Intensif (C\'ECI), funded by the Fonds de la Recherche Scientifique (F.R.S.-FNRS) under Grant No. 2.5020.11.
MF \& EB acknowledges FNRS for support and the PDR project CHRYSALID No.40003544. Work at West Virginia University was supported by the U.S. Department of Energy (DOE), Office of Science, Basic Energy Sciences (BES) under Award DE‐SC0021375. We also acknowledge the computational
resources awarded by XSEDE, a project supported by National Science
Foundation grant number ACI-1053575. The authors also acknowledge the
support from the Texas Advances Computer Center (with the Stampede2
and Bridges supercomputers). We also acknowledge the Super Computing
System (Thorny Flat) at WVU, which is funded in part by the National
Science Foundation (NSF) Major Research Instrumentation Program (MRI)
Award \#1726534, and West Virginia University.


\bibliographystyle{apsrev4-2}
\bibliography{biblio}

\begin{thebibliography}{62}%
\makeatletter
\providecommand \@ifxundefined [1]{%
 \@ifx{#1\undefined}
}%
\providecommand \@ifnum [1]{%
 \ifnum #1\expandafter \@firstoftwo
 \else \expandafter \@secondoftwo
 \fi
}%
\providecommand \@ifx [1]{%
 \ifx #1\expandafter \@firstoftwo
 \else \expandafter \@secondoftwo
 \fi
}%
\providecommand \natexlab [1]{#1}%
\providecommand \enquote  [1]{``#1''}%
\providecommand \bibnamefont  [1]{#1}%
\providecommand \bibfnamefont [1]{#1}%
\providecommand \citenamefont [1]{#1}%
\providecommand \href@noop [0]{\@secondoftwo}%
\providecommand \href [0]{\begingroup \@sanitize@url \@href}%
\providecommand \@href[1]{\@@startlink{#1}\@@href}%
\providecommand \@@href[1]{\endgroup#1\@@endlink}%
\providecommand \@sanitize@url [0]{\catcode `\\12\catcode `\$12\catcode
  `\&12\catcode `\#12\catcode `\^12\catcode `\_12\catcode `\%12\relax}%
\providecommand \@@startlink[1]{}%
\providecommand \@@endlink[0]{}%
\providecommand \url  [0]{\begingroup\@sanitize@url \@url }%
\providecommand \@url [1]{\endgroup\@href {#1}{\urlprefix }}%
\providecommand \urlprefix  [0]{URL }%
\providecommand \Eprint [0]{\href }%
\providecommand \doibase [0]{https://doi.org/}%
\providecommand \selectlanguage [0]{\@gobble}%
\providecommand \bibinfo  [0]{\@secondoftwo}%
\providecommand \bibfield  [0]{\@secondoftwo}%
\providecommand \translation [1]{[#1]}%
\providecommand \BibitemOpen [0]{}%
\providecommand \bibitemStop [0]{}%
\providecommand \bibitemNoStop [0]{.\EOS\space}%
\providecommand \EOS [0]{\spacefactor3000\relax}%
\providecommand \BibitemShut  [1]{\csname bibitem#1\endcsname}%
\let\auto@bib@innerbib\@empty
\bibitem [{\citenamefont {Stöhr}\ and\ \citenamefont
  {Siegmann}(2006)}]{stohr2006}%
  \BibitemOpen
  \bibfield  {author} {\bibinfo {author} {\bibfnamefont {J.}~\bibnamefont
  {Stöhr}}\ and\ \bibinfo {author} {\bibfnamefont {H.~C.}\ \bibnamefont
  {Siegmann}},\ }\href {https://doi.org/10.1007/978-3-540-30283-4} {\emph
  {\bibinfo {title} {Magnetism: From Fundamentals to Nanoscale Dynamics}}}\
  (\bibinfo  {publisher} {Springer Berlin Heidelberg},\ \bibinfo {address}
  {Berlin, Heidelberg},\ \bibinfo {year} {2006})\BibitemShut {NoStop}%
\bibitem [{\citenamefont {Manchon}\ \emph {et~al.}(2019)\citenamefont
  {Manchon}, \citenamefont {\ifmmode~\check{Z}\else \v{Z}\fi{}elezn\'y},
  \citenamefont {Miron}, \citenamefont {Jungwirth}, \citenamefont {Sinova},
  \citenamefont {Thiaville}, \citenamefont {Garello},\ and\ \citenamefont
  {Gambardella}}]{RevModPhys.91.035004}%
  \BibitemOpen
  \bibfield  {author} {\bibinfo {author} {\bibfnamefont {A.}~\bibnamefont
  {Manchon}}, \bibinfo {author} {\bibfnamefont {J.}~\bibnamefont
  {\ifmmode~\check{Z}\else \v{Z}\fi{}elezn\'y}}, \bibinfo {author}
  {\bibfnamefont {I.~M.}\ \bibnamefont {Miron}}, \bibinfo {author}
  {\bibfnamefont {T.}~\bibnamefont {Jungwirth}}, \bibinfo {author}
  {\bibfnamefont {J.}~\bibnamefont {Sinova}}, \bibinfo {author} {\bibfnamefont
  {A.}~\bibnamefont {Thiaville}}, \bibinfo {author} {\bibfnamefont
  {K.}~\bibnamefont {Garello}},\ and\ \bibinfo {author} {\bibfnamefont
  {P.}~\bibnamefont {Gambardella}},\ }\href
  {https://doi.org/10.1103/RevModPhys.91.035004} {\bibfield  {journal}
  {\bibinfo  {journal} {Rev. Mod. Phys.}\ }\textbf {\bibinfo {volume} {91}},\
  \bibinfo {pages} {035004} (\bibinfo {year} {2019})}\BibitemShut {NoStop}%
\bibitem [{\citenamefont {Gambardella}\ and\ \citenamefont
  {Miron}(2011)}]{torque_2011}%
  \BibitemOpen
  \bibfield  {author} {\bibinfo {author} {\bibfnamefont {P.}~\bibnamefont
  {Gambardella}}\ and\ \bibinfo {author} {\bibfnamefont {I.~M.}\ \bibnamefont
  {Miron}},\ }\href {https://doi.org/10.1098/rsta.2010.0336} {\bibfield
  {journal} {\bibinfo  {journal} {Phil. Trans. R. Soc. A}\ }\textbf {\bibinfo
  {volume} {369}},\ \bibinfo {pages} {3175} (\bibinfo {year}
  {2011})}\BibitemShut {NoStop}%
\bibitem [{\citenamefont {Tokura}\ and\ \citenamefont
  {Kanazawa}(2021)}]{tokura2021}%
  \BibitemOpen
  \bibfield  {author} {\bibinfo {author} {\bibfnamefont {Y.}~\bibnamefont
  {Tokura}}\ and\ \bibinfo {author} {\bibfnamefont {N.}~\bibnamefont
  {Kanazawa}},\ }\href {https://doi.org/10.1021/acs.chemrev.0c00297} {\bibfield
   {journal} {\bibinfo  {journal} {Chem. Rev.}\ }\textbf {\bibinfo {volume}
  {121}},\ \bibinfo {pages} {2857} (\bibinfo {year} {2021})}\BibitemShut
  {NoStop}%
\bibitem [{\citenamefont {Everschor-Sitte}\ \emph {et~al.}(2018)\citenamefont
  {Everschor-Sitte}, \citenamefont {Masell}, \citenamefont {Reeve},\ and\
  \citenamefont {Kl{\"a}ui}}]{everschor-sitte2018}%
  \BibitemOpen
  \bibfield  {author} {\bibinfo {author} {\bibfnamefont {K.}~\bibnamefont
  {Everschor-Sitte}}, \bibinfo {author} {\bibfnamefont {J.}~\bibnamefont
  {Masell}}, \bibinfo {author} {\bibfnamefont {R.~M.}\ \bibnamefont {Reeve}},\
  and\ \bibinfo {author} {\bibfnamefont {M.}~\bibnamefont {Kl{\"a}ui}},\ }\href
  {https://doi.org/10.1063/1.5048972} {\bibfield  {journal} {\bibinfo
  {journal} {J. Appl. Phys.}\ }\textbf {\bibinfo {volume} {124}},\ \bibinfo
  {pages} {240901} (\bibinfo {year} {2018})}\BibitemShut {NoStop}%
\bibitem [{\citenamefont {Fert}\ \emph {et~al.}(2017)\citenamefont {Fert},
  \citenamefont {Reyren},\ and\ \citenamefont {Cros}}]{fert2017}%
  \BibitemOpen
  \bibfield  {author} {\bibinfo {author} {\bibfnamefont {A.}~\bibnamefont
  {Fert}}, \bibinfo {author} {\bibfnamefont {N.}~\bibnamefont {Reyren}},\ and\
  \bibinfo {author} {\bibfnamefont {V.}~\bibnamefont {Cros}},\ }\href
  {https://doi.org/10.1038/natrevmats.2017.31} {\bibfield  {journal} {\bibinfo
  {journal} {Nat. Rev. Mater.}\ }\textbf {\bibinfo {volume} {2}},\ \bibinfo
  {pages} {17031} (\bibinfo {year} {2017})}\BibitemShut {NoStop}%
\bibitem [{\citenamefont {Hasan}\ and\ \citenamefont
  {Kane}(2010)}]{topo_review}%
  \BibitemOpen
  \bibfield  {author} {\bibinfo {author} {\bibfnamefont {M.~Z.}\ \bibnamefont
  {Hasan}}\ and\ \bibinfo {author} {\bibfnamefont {C.~L.}\ \bibnamefont
  {Kane}},\ }\href {https://doi.org/10.1103/RevModPhys.82.3045} {\bibfield
  {journal} {\bibinfo  {journal} {Rev. Mod. Phys.}\ }\textbf {\bibinfo {volume}
  {82}},\ \bibinfo {pages} {3045} (\bibinfo {year} {2010})}\BibitemShut
  {NoStop}%
\bibitem [{\citenamefont {Qi}\ and\ \citenamefont
  {Zhang}(2011)}]{topo_review_2}%
  \BibitemOpen
  \bibfield  {author} {\bibinfo {author} {\bibfnamefont {X.-L.}\ \bibnamefont
  {Qi}}\ and\ \bibinfo {author} {\bibfnamefont {S.-C.}\ \bibnamefont {Zhang}},\
  }\href {https://doi.org/10.1103/RevModPhys.83.1057} {\bibfield  {journal}
  {\bibinfo  {journal} {Rev. Mod. Phys.}\ }\textbf {\bibinfo {volume} {83}},\
  \bibinfo {pages} {1057} (\bibinfo {year} {2011})}\BibitemShut {NoStop}%
\bibitem [{\citenamefont {Murakami}\ \emph {et~al.}(2003)\citenamefont
  {Murakami}, \citenamefont {Nagaosa},\ and\ \citenamefont
  {Zhang}}]{Murakami2003}%
  \BibitemOpen
  \bibfield  {author} {\bibinfo {author} {\bibfnamefont {S.}~\bibnamefont
  {Murakami}}, \bibinfo {author} {\bibfnamefont {N.}~\bibnamefont {Nagaosa}},\
  and\ \bibinfo {author} {\bibfnamefont {S.-C.}\ \bibnamefont {Zhang}},\ }\href
  {https://doi.org/10.1126/science.1087128} {\bibfield  {journal} {\bibinfo
  {journal} {Science}\ }\textbf {\bibinfo {volume} {301}},\ \bibinfo {pages}
  {1348} (\bibinfo {year} {2003})}\BibitemShut {NoStop}%
\bibitem [{\citenamefont {Sinova}\ \emph {et~al.}(2004)\citenamefont {Sinova},
  \citenamefont {Culcer}, \citenamefont {Niu}, \citenamefont {Sinitsyn},
  \citenamefont {Jungwirth},\ and\ \citenamefont {MacDonald}}]{Sinova}%
  \BibitemOpen
  \bibfield  {author} {\bibinfo {author} {\bibfnamefont {J.}~\bibnamefont
  {Sinova}}, \bibinfo {author} {\bibfnamefont {D.}~\bibnamefont {Culcer}},
  \bibinfo {author} {\bibfnamefont {Q.}~\bibnamefont {Niu}}, \bibinfo {author}
  {\bibfnamefont {N.~A.}\ \bibnamefont {Sinitsyn}}, \bibinfo {author}
  {\bibfnamefont {T.}~\bibnamefont {Jungwirth}},\ and\ \bibinfo {author}
  {\bibfnamefont {A.~H.}\ \bibnamefont {MacDonald}},\ }\href
  {https://doi.org/10.1103/PhysRevLett.92.126603} {\bibfield  {journal}
  {\bibinfo  {journal} {Phys. Rev. Lett.}\ }\textbf {\bibinfo {volume} {92}},\
  \bibinfo {pages} {126603} (\bibinfo {year} {2004})}\BibitemShut {NoStop}%
\bibitem [{\citenamefont {Kato}\ \emph {et~al.}(2004)\citenamefont {Kato},
  \citenamefont {Myers}, \citenamefont {Gossard},\ and\ \citenamefont
  {Awschalom}}]{Kato2004}%
  \BibitemOpen
  \bibfield  {author} {\bibinfo {author} {\bibfnamefont {Y.~K.}\ \bibnamefont
  {Kato}}, \bibinfo {author} {\bibfnamefont {R.~C.}\ \bibnamefont {Myers}},
  \bibinfo {author} {\bibfnamefont {A.~C.}\ \bibnamefont {Gossard}},\ and\
  \bibinfo {author} {\bibfnamefont {D.~D.}\ \bibnamefont {Awschalom}},\ }\href
  {https://doi.org/10.1126/science.1105514} {\bibfield  {journal} {\bibinfo
  {journal} {Science}\ }\textbf {\bibinfo {volume} {306}},\ \bibinfo {pages}
  {1910} (\bibinfo {year} {2004})}\BibitemShut {NoStop}%
\bibitem [{\citenamefont {Bernevig}\ and\ \citenamefont
  {Zhang}(2006)}]{QSHE_Bernevig}%
  \BibitemOpen
  \bibfield  {author} {\bibinfo {author} {\bibfnamefont {B.~A.}\ \bibnamefont
  {Bernevig}}\ and\ \bibinfo {author} {\bibfnamefont {S.-C.}\ \bibnamefont
  {Zhang}},\ }\href {https://doi.org/10.1103/PhysRevLett.96.106802} {\bibfield
  {journal} {\bibinfo  {journal} {Phys. Rev. Lett.}\ }\textbf {\bibinfo
  {volume} {96}},\ \bibinfo {pages} {106802} (\bibinfo {year}
  {2006})}\BibitemShut {NoStop}%
\bibitem [{\citenamefont {Koralek}\ \emph {et~al.}(2009)\citenamefont
  {Koralek}, \citenamefont {Weber}, \citenamefont {Orenstein}, \citenamefont
  {Bernevig}, \citenamefont {Zhang}, \citenamefont {Mack},\ and\ \citenamefont
  {Awschalom}}]{Koralek2009}%
  \BibitemOpen
  \bibfield  {author} {\bibinfo {author} {\bibfnamefont {J.~D.}\ \bibnamefont
  {Koralek}}, \bibinfo {author} {\bibfnamefont {C.~P.}\ \bibnamefont {Weber}},
  \bibinfo {author} {\bibfnamefont {J.}~\bibnamefont {Orenstein}}, \bibinfo
  {author} {\bibfnamefont {B.~A.}\ \bibnamefont {Bernevig}}, \bibinfo {author}
  {\bibfnamefont {S.-C.}\ \bibnamefont {Zhang}}, \bibinfo {author}
  {\bibfnamefont {S.}~\bibnamefont {Mack}},\ and\ \bibinfo {author}
  {\bibfnamefont {D.~D.}\ \bibnamefont {Awschalom}},\ }\href
  {https://doi.org/10.1038/nature07871} {\bibfield  {journal} {\bibinfo
  {journal} {Nature}\ }\textbf {\bibinfo {volume} {458}},\ \bibinfo {pages}
  {610} (\bibinfo {year} {2009})}\BibitemShut {NoStop}%
\bibitem [{\citenamefont {Walser}\ \emph {et~al.}(2012)\citenamefont {Walser},
  \citenamefont {Reichl}, \citenamefont {Wegscheider},\ and\ \citenamefont
  {Salis}}]{Walser2012}%
  \BibitemOpen
  \bibfield  {author} {\bibinfo {author} {\bibfnamefont {M.~P.}\ \bibnamefont
  {Walser}}, \bibinfo {author} {\bibfnamefont {C.}~\bibnamefont {Reichl}},
  \bibinfo {author} {\bibfnamefont {W.}~\bibnamefont {Wegscheider}},\ and\
  \bibinfo {author} {\bibfnamefont {G.}~\bibnamefont {Salis}},\ }\href
  {https://doi.org/10.1038/nphys2383} {\bibfield  {journal} {\bibinfo
  {journal} {Nat. Phys.}\ }\textbf {\bibinfo {volume} {8}},\ \bibinfo {pages}
  {757} (\bibinfo {year} {2012})}\BibitemShut {NoStop}%
\bibitem [{\citenamefont {Sasaki}\ \emph {et~al.}(2014)\citenamefont {Sasaki},
  \citenamefont {Nonaka}, \citenamefont {Kunihashi}, \citenamefont {Kohda},
  \citenamefont {Bauernfeind}, \citenamefont {Dollinger}, \citenamefont
  {Richter},\ and\ \citenamefont {Nitta}}]{Sasaki2014}%
  \BibitemOpen
  \bibfield  {author} {\bibinfo {author} {\bibfnamefont {A.}~\bibnamefont
  {Sasaki}}, \bibinfo {author} {\bibfnamefont {S.}~\bibnamefont {Nonaka}},
  \bibinfo {author} {\bibfnamefont {Y.}~\bibnamefont {Kunihashi}}, \bibinfo
  {author} {\bibfnamefont {M.}~\bibnamefont {Kohda}}, \bibinfo {author}
  {\bibfnamefont {T.}~\bibnamefont {Bauernfeind}}, \bibinfo {author}
  {\bibfnamefont {T.}~\bibnamefont {Dollinger}}, \bibinfo {author}
  {\bibfnamefont {K.}~\bibnamefont {Richter}},\ and\ \bibinfo {author}
  {\bibfnamefont {J.}~\bibnamefont {Nitta}},\ }\href
  {https://doi.org/10.1038/nnano.2014.128} {\bibfield  {journal} {\bibinfo
  {journal} {Nat. Nanotechnol.}\ }\textbf {\bibinfo {volume} {9}},\ \bibinfo
  {pages} {703} (\bibinfo {year} {2014})}\BibitemShut {NoStop}%
\bibitem [{\citenamefont {Tao}\ and\ \citenamefont {Tsymbal}(2018)}]{Tao2018}%
  \BibitemOpen
  \bibfield  {author} {\bibinfo {author} {\bibfnamefont {L.~L.}\ \bibnamefont
  {Tao}}\ and\ \bibinfo {author} {\bibfnamefont {E.~Y.}\ \bibnamefont
  {Tsymbal}},\ }\href {https://doi.org/10.1038/s41467-018-05137-0} {\bibfield
  {journal} {\bibinfo  {journal} {Nat. Commun.}\ }\textbf {\bibinfo {volume}
  {9}},\ \bibinfo {pages} {2763} (\bibinfo {year} {2018})}\BibitemShut
  {NoStop}%
\bibitem [{\citenamefont {Tao}\ and\ \citenamefont {Tsymbal}(2021)}]{Tao_2021}%
  \BibitemOpen
  \bibfield  {author} {\bibinfo {author} {\bibfnamefont {L.~L.}\ \bibnamefont
  {Tao}}\ and\ \bibinfo {author} {\bibfnamefont {E.~Y.}\ \bibnamefont
  {Tsymbal}},\ }\href {https://doi.org/10.1088/1361-6463/abcc25} {\bibfield
  {journal} {\bibinfo  {journal} {J. Phys. D: Appl. Phys.}\ }\textbf {\bibinfo
  {volume} {54}},\ \bibinfo {pages} {113001} (\bibinfo {year}
  {2021})}\BibitemShut {NoStop}%
\bibitem [{\citenamefont {Rashba}(1960)}]{Rashba_1960}%
  \BibitemOpen
  \bibfield  {author} {\bibinfo {author} {\bibfnamefont {E.}~\bibnamefont
  {Rashba}},\ }\href {https://cir.nii.ac.jp/crid/1571698600346713472}
  {\bibfield  {journal} {\bibinfo  {journal} {Sov. Phys.-Solid State}\ }\textbf
  {\bibinfo {volume} {2}},\ \bibinfo {pages} {1109} (\bibinfo {year}
  {1960})}\BibitemShut {NoStop}%
\bibitem [{\citenamefont {Bychkov}\ and\ \citenamefont
  {Rashba}(1984)}]{Yu_Bychkov_1984}%
  \BibitemOpen
  \bibfield  {author} {\bibinfo {author} {\bibfnamefont {Y.~A.}\ \bibnamefont
  {Bychkov}}\ and\ \bibinfo {author} {\bibfnamefont {E.~I.}\ \bibnamefont
  {Rashba}},\ }\href {https://doi.org/10.1088/0022-3719/17/33/015} {\bibfield
  {journal} {\bibinfo  {journal} {J. Phys. C: Solid State Phys.}\ }\textbf
  {\bibinfo {volume} {17}},\ \bibinfo {pages} {6039} (\bibinfo {year}
  {1984})}\BibitemShut {NoStop}%
\bibitem [{\citenamefont {Di~Sante}\ \emph {et~al.}(2013)\citenamefont
  {Di~Sante}, \citenamefont {Barone}, \citenamefont {Bertacco},\ and\
  \citenamefont {Picozzi}}]{di_Sante_Rashba_2013}%
  \BibitemOpen
  \bibfield  {author} {\bibinfo {author} {\bibfnamefont {D.}~\bibnamefont
  {Di~Sante}}, \bibinfo {author} {\bibfnamefont {P.}~\bibnamefont {Barone}},
  \bibinfo {author} {\bibfnamefont {R.}~\bibnamefont {Bertacco}},\ and\
  \bibinfo {author} {\bibfnamefont {S.}~\bibnamefont {Picozzi}},\ }\href
  {https://doi.org/https://doi.org/10.1002/adma.201203199} {\bibfield
  {journal} {\bibinfo  {journal} {Adv. Mater.}\ }\textbf {\bibinfo {volume}
  {25}},\ \bibinfo {pages} {509} (\bibinfo {year} {2013})}\BibitemShut
  {NoStop}%
\bibitem [{\citenamefont {Tao}\ \emph {et~al.}(2017)\citenamefont {Tao},
  \citenamefont {Paudel}, \citenamefont {Kovalev},\ and\ \citenamefont
  {Tsymbal}}]{PhysRevB.95.245141}%
  \BibitemOpen
  \bibfield  {author} {\bibinfo {author} {\bibfnamefont {L.~L.}\ \bibnamefont
  {Tao}}, \bibinfo {author} {\bibfnamefont {T.~R.}\ \bibnamefont {Paudel}},
  \bibinfo {author} {\bibfnamefont {A.~A.}\ \bibnamefont {Kovalev}},\ and\
  \bibinfo {author} {\bibfnamefont {E.~Y.}\ \bibnamefont {Tsymbal}},\ }\href
  {https://doi.org/10.1103/PhysRevB.95.245141} {\bibfield  {journal} {\bibinfo
  {journal} {Phys. Rev. B}\ }\textbf {\bibinfo {volume} {95}},\ \bibinfo
  {pages} {245141} (\bibinfo {year} {2017})}\BibitemShut {NoStop}%
\bibitem [{\citenamefont {Dresselhaus}(1955)}]{PhysRev.100.580}%
  \BibitemOpen
  \bibfield  {author} {\bibinfo {author} {\bibfnamefont {G.}~\bibnamefont
  {Dresselhaus}},\ }\href {https://doi.org/10.1103/PhysRev.100.580} {\bibfield
  {journal} {\bibinfo  {journal} {Phys. Rev.}\ }\textbf {\bibinfo {volume}
  {100}},\ \bibinfo {pages} {580} (\bibinfo {year} {1955})}\BibitemShut
  {NoStop}%
\bibitem [{\citenamefont {Moriya}\ \emph {et~al.}(2014)\citenamefont {Moriya},
  \citenamefont {Sawano}, \citenamefont {Hoshi}, \citenamefont {Masubuchi},
  \citenamefont {Shiraki}, \citenamefont {Wild}, \citenamefont {Neumann},
  \citenamefont {Abstreiter}, \citenamefont {Bougeard}, \citenamefont {Koga},\
  and\ \citenamefont {Machida}}]{PhysRevLett.113.086601}%
  \BibitemOpen
  \bibfield  {author} {\bibinfo {author} {\bibfnamefont {R.}~\bibnamefont
  {Moriya}}, \bibinfo {author} {\bibfnamefont {K.}~\bibnamefont {Sawano}},
  \bibinfo {author} {\bibfnamefont {Y.}~\bibnamefont {Hoshi}}, \bibinfo
  {author} {\bibfnamefont {S.}~\bibnamefont {Masubuchi}}, \bibinfo {author}
  {\bibfnamefont {Y.}~\bibnamefont {Shiraki}}, \bibinfo {author} {\bibfnamefont
  {A.}~\bibnamefont {Wild}}, \bibinfo {author} {\bibfnamefont {C.}~\bibnamefont
  {Neumann}}, \bibinfo {author} {\bibfnamefont {G.}~\bibnamefont {Abstreiter}},
  \bibinfo {author} {\bibfnamefont {D.}~\bibnamefont {Bougeard}}, \bibinfo
  {author} {\bibfnamefont {T.}~\bibnamefont {Koga}},\ and\ \bibinfo {author}
  {\bibfnamefont {T.}~\bibnamefont {Machida}},\ }\href
  {https://doi.org/10.1103/PhysRevLett.113.086601} {\bibfield  {journal}
  {\bibinfo  {journal} {Phys. Rev. Lett.}\ }\textbf {\bibinfo {volume} {113}},\
  \bibinfo {pages} {086601} (\bibinfo {year} {2014})}\BibitemShut {NoStop}%
\bibitem [{\citenamefont {Nakamura}\ \emph {et~al.}(2012)\citenamefont
  {Nakamura}, \citenamefont {Koga},\ and\ \citenamefont
  {Kimura}}]{PhysRevLett.108.206601}%
  \BibitemOpen
  \bibfield  {author} {\bibinfo {author} {\bibfnamefont {H.}~\bibnamefont
  {Nakamura}}, \bibinfo {author} {\bibfnamefont {T.}~\bibnamefont {Koga}},\
  and\ \bibinfo {author} {\bibfnamefont {T.}~\bibnamefont {Kimura}},\ }\href
  {https://doi.org/10.1103/PhysRevLett.108.206601} {\bibfield  {journal}
  {\bibinfo  {journal} {Phys. Rev. Lett.}\ }\textbf {\bibinfo {volume} {108}},\
  \bibinfo {pages} {206601} (\bibinfo {year} {2012})}\BibitemShut {NoStop}%
\bibitem [{\citenamefont {Gmitra}\ and\ \citenamefont
  {Fabian}(2016)}]{PhysRevB.94.165202}%
  \BibitemOpen
  \bibfield  {author} {\bibinfo {author} {\bibfnamefont {M.}~\bibnamefont
  {Gmitra}}\ and\ \bibinfo {author} {\bibfnamefont {J.}~\bibnamefont
  {Fabian}},\ }\href {https://doi.org/10.1103/PhysRevB.94.165202} {\bibfield
  {journal} {\bibinfo  {journal} {Phys. Rev. B}\ }\textbf {\bibinfo {volume}
  {94}},\ \bibinfo {pages} {165202} (\bibinfo {year} {2016})}\BibitemShut
  {NoStop}%
\bibitem [{\citenamefont {Trier}\ \emph {et~al.}(2022)\citenamefont {Trier},
  \citenamefont {No{\"e}l}, \citenamefont {Kim}, \citenamefont {Attan{\'e}},
  \citenamefont {Vila},\ and\ \citenamefont {Bibes}}]{trier2022}%
  \BibitemOpen
  \bibfield  {author} {\bibinfo {author} {\bibfnamefont {F.}~\bibnamefont
  {Trier}}, \bibinfo {author} {\bibfnamefont {P.}~\bibnamefont {No{\"e}l}},
  \bibinfo {author} {\bibfnamefont {J.-V.}\ \bibnamefont {Kim}}, \bibinfo
  {author} {\bibfnamefont {J.-P.}\ \bibnamefont {Attan{\'e}}}, \bibinfo
  {author} {\bibfnamefont {L.}~\bibnamefont {Vila}},\ and\ \bibinfo {author}
  {\bibfnamefont {M.}~\bibnamefont {Bibes}},\ }\href
  {https://doi.org/10.1038/s41578-021-00395-9} {\bibfield  {journal} {\bibinfo
  {journal} {Nat. Rev. Mater.}\ }\textbf {\bibinfo {volume} {7}},\ \bibinfo
  {pages} {258} (\bibinfo {year} {2022})}\BibitemShut {NoStop}%
\bibitem [{\citenamefont {Djani}\ \emph {et~al.}(2019)\citenamefont {Djani},
  \citenamefont {Garcia-Castro}, \citenamefont {Tong}, \citenamefont {Barone},
  \citenamefont {Bousquet}, \citenamefont {Picozzi},\ and\ \citenamefont
  {Ghosez}}]{Djani2019}%
  \BibitemOpen
  \bibfield  {author} {\bibinfo {author} {\bibfnamefont {H.}~\bibnamefont
  {Djani}}, \bibinfo {author} {\bibfnamefont {A.~C.}\ \bibnamefont
  {Garcia-Castro}}, \bibinfo {author} {\bibfnamefont {W.-Y.}\ \bibnamefont
  {Tong}}, \bibinfo {author} {\bibfnamefont {P.}~\bibnamefont {Barone}},
  \bibinfo {author} {\bibfnamefont {E.}~\bibnamefont {Bousquet}}, \bibinfo
  {author} {\bibfnamefont {S.}~\bibnamefont {Picozzi}},\ and\ \bibinfo {author}
  {\bibfnamefont {P.}~\bibnamefont {Ghosez}},\ }\href
  {https://doi.org/10.1038/s41535-019-0190-z} {\bibfield  {journal} {\bibinfo
  {journal} {npj Quantum Materials}\ }\textbf {\bibinfo {volume} {4}},\
  \bibinfo {pages} {51} (\bibinfo {year} {2019})}\BibitemShut {NoStop}%
\bibitem [{\citenamefont {Iwasaki}\ \emph {et~al.}(1971)\citenamefont
  {Iwasaki}, \citenamefont {Sugii}, \citenamefont {Yamada},\ and\ \citenamefont
  {Niizeki}}]{iwasaki1971}%
  \BibitemOpen
  \bibfield  {author} {\bibinfo {author} {\bibfnamefont {H.}~\bibnamefont
  {Iwasaki}}, \bibinfo {author} {\bibfnamefont {K.}~\bibnamefont {Sugii}},
  \bibinfo {author} {\bibfnamefont {T.}~\bibnamefont {Yamada}},\ and\ \bibinfo
  {author} {\bibfnamefont {N.}~\bibnamefont {Niizeki}},\ }\href
  {https://doi.org/10.1063/1.1653487} {\bibfield  {journal} {\bibinfo
  {journal} {Appl. Phys. Lett.}\ }\textbf {\bibinfo {volume} {18}},\ \bibinfo
  {pages} {444} (\bibinfo {year} {1971})}\BibitemShut {NoStop}%
\bibitem [{\citenamefont {Iwasaki}\ \emph {et~al.}(1972)\citenamefont
  {Iwasaki}, \citenamefont {Miyazawa}, \citenamefont {Koizumi}, \citenamefont
  {Sugii},\ and\ \citenamefont {Niizeki}}]{Iwasaki1972}%
  \BibitemOpen
  \bibfield  {author} {\bibinfo {author} {\bibfnamefont {H.}~\bibnamefont
  {Iwasaki}}, \bibinfo {author} {\bibfnamefont {S.}~\bibnamefont {Miyazawa}},
  \bibinfo {author} {\bibfnamefont {H.}~\bibnamefont {Koizumi}}, \bibinfo
  {author} {\bibfnamefont {K.}~\bibnamefont {Sugii}},\ and\ \bibinfo {author}
  {\bibfnamefont {N.}~\bibnamefont {Niizeki}},\ }\href@noop {} {\bibfield
  {journal} {\bibinfo  {journal} {J. Appl. Phys.}\ }\textbf {\bibinfo {volume}
  {43}},\ \bibinfo {pages} {4907} (\bibinfo {year} {1972})}\BibitemShut
  {NoStop}%
\bibitem [{\citenamefont {Gonze}\ and\ \citenamefont
  {et~al.}(2020)}]{gonze2020}%
  \BibitemOpen
  \bibfield  {author} {\bibinfo {author} {\bibfnamefont {X.}~\bibnamefont
  {Gonze}}\ and\ \bibinfo {author} {\bibnamefont {et~al.}},\ }\href
  {https://doi.org/https://doi.org/10.1016/j.cpc.2019.107042} {\bibfield
  {journal} {\bibinfo  {journal} {Computer Physics Communications}\ }\textbf
  {\bibinfo {volume} {248}},\ \bibinfo {pages} {107042} (\bibinfo {year}
  {2020})}\BibitemShut {NoStop}%
\bibitem [{\citenamefont {Kresse}\ and\ \citenamefont
  {Joubert}(1999)}]{Kresse1999}%
  \BibitemOpen
  \bibfield  {author} {\bibinfo {author} {\bibfnamefont {G.}~\bibnamefont
  {Kresse}}\ and\ \bibinfo {author} {\bibfnamefont {D.}~\bibnamefont
  {Joubert}},\ }\href {https://doi.org/10.1103/PhysRevB.59.1758} {\bibfield
  {journal} {\bibinfo  {journal} {Phys. Rev. B}\ }\textbf {\bibinfo {volume}
  {59}},\ \bibinfo {pages} {1758} (\bibinfo {year} {1999})}\BibitemShut
  {NoStop}%
\bibitem [{\citenamefont {Konak}\ \emph {et~al.}(1978)\citenamefont {Konak},
  \citenamefont {Kopsky},\ and\ \citenamefont {Smutny}}]{C_Konak_1978}%
  \BibitemOpen
  \bibfield  {author} {\bibinfo {author} {\bibfnamefont {C.}~\bibnamefont
  {Konak}}, \bibinfo {author} {\bibfnamefont {V.}~\bibnamefont {Kopsky}},\ and\
  \bibinfo {author} {\bibfnamefont {F.}~\bibnamefont {Smutny}},\ }\href
  {https://doi.org/10.1088/0022-3719/11/12/012} {\bibfield  {journal} {\bibinfo
   {journal} {J. Phys. C: Solid State Phys.}\ }\textbf {\bibinfo {volume}
  {11}},\ \bibinfo {pages} {2493} (\bibinfo {year} {1978})}\BibitemShut
  {NoStop}%
\bibitem [{\citenamefont {Vlokh}(1987)}]{Vlokh_1987}%
  \BibitemOpen
  \bibfield  {author} {\bibinfo {author} {\bibfnamefont {O.~G.}\ \bibnamefont
  {Vlokh}},\ }\href {https://doi.org/10.1080/00150198708008216} {\bibfield
  {journal} {\bibinfo  {journal} {Ferroelectrics}\ }\textbf {\bibinfo {volume}
  {75}},\ \bibinfo {pages} {119} (\bibinfo {year} {1987})}\BibitemShut
  {NoStop}%
\bibitem [{\citenamefont {Arras}\ \emph {et~al.}(2019)\citenamefont {Arras},
  \citenamefont {Gosteau}, \citenamefont {Zhao}, \citenamefont {Paillard},
  \citenamefont {Yang},\ and\ \citenamefont {Bellaiche}}]{arras2019}%
  \BibitemOpen
  \bibfield  {author} {\bibinfo {author} {\bibfnamefont {R.}~\bibnamefont
  {Arras}}, \bibinfo {author} {\bibfnamefont {J.}~\bibnamefont {Gosteau}},
  \bibinfo {author} {\bibfnamefont {H.~J.}\ \bibnamefont {Zhao}}, \bibinfo
  {author} {\bibfnamefont {C.}~\bibnamefont {Paillard}}, \bibinfo {author}
  {\bibfnamefont {Y.}~\bibnamefont {Yang}},\ and\ \bibinfo {author}
  {\bibfnamefont {L.}~\bibnamefont {Bellaiche}},\ }\href
  {https://doi.org/10.1103/PhysRevB.100.174415} {\bibfield  {journal} {\bibinfo
   {journal} {Phys. Rev. B}\ }\textbf {\bibinfo {volume} {100}},\ \bibinfo
  {pages} {174415} (\bibinfo {year} {2019})}\BibitemShut {NoStop}%
\bibitem [{\citenamefont {Rangel}\ \emph {et~al.}(2012)\citenamefont {Rangel},
  \citenamefont {Kecik}, \citenamefont {Trevisanutto}, \citenamefont
  {Rignanese}, \citenamefont {Van~Swygenhoven},\ and\ \citenamefont
  {Olevano}}]{rangel2012}%
  \BibitemOpen
  \bibfield  {author} {\bibinfo {author} {\bibfnamefont {T.}~\bibnamefont
  {Rangel}}, \bibinfo {author} {\bibfnamefont {D.}~\bibnamefont {Kecik}},
  \bibinfo {author} {\bibfnamefont {P.~E.}\ \bibnamefont {Trevisanutto}},
  \bibinfo {author} {\bibfnamefont {G.-M.}\ \bibnamefont {Rignanese}}, \bibinfo
  {author} {\bibfnamefont {H.}~\bibnamefont {Van~Swygenhoven}},\ and\ \bibinfo
  {author} {\bibfnamefont {V.}~\bibnamefont {Olevano}},\ }\href
  {https://doi.org/10.1103/PhysRevB.86.125125} {\bibfield  {journal} {\bibinfo
  {journal} {Phys. Rev. B}\ }\textbf {\bibinfo {volume} {86}},\ \bibinfo
  {pages} {125125} (\bibinfo {year} {2012})}\BibitemShut {NoStop}%
\bibitem [{\citenamefont {Zhao}\ \emph {et~al.}(2020)\citenamefont {Zhao},
  \citenamefont {Nakamura}, \citenamefont {Arras}, \citenamefont {Paillard},
  \citenamefont {Chen}, \citenamefont {Gosteau}, \citenamefont {Li},
  \citenamefont {Yang},\ and\ \citenamefont
  {Bellaiche}}]{PhysRevLett.125.216405}%
  \BibitemOpen
  \bibfield  {author} {\bibinfo {author} {\bibfnamefont {H.~J.}\ \bibnamefont
  {Zhao}}, \bibinfo {author} {\bibfnamefont {H.}~\bibnamefont {Nakamura}},
  \bibinfo {author} {\bibfnamefont {R.}~\bibnamefont {Arras}}, \bibinfo
  {author} {\bibfnamefont {C.}~\bibnamefont {Paillard}}, \bibinfo {author}
  {\bibfnamefont {P.}~\bibnamefont {Chen}}, \bibinfo {author} {\bibfnamefont
  {J.}~\bibnamefont {Gosteau}}, \bibinfo {author} {\bibfnamefont
  {X.}~\bibnamefont {Li}}, \bibinfo {author} {\bibfnamefont {Y.}~\bibnamefont
  {Yang}},\ and\ \bibinfo {author} {\bibfnamefont {L.}~\bibnamefont
  {Bellaiche}},\ }\href {https://doi.org/10.1103/PhysRevLett.125.216405}
  {\bibfield  {journal} {\bibinfo  {journal} {Phys. Rev. Lett.}\ }\textbf
  {\bibinfo {volume} {125}},\ \bibinfo {pages} {216405} (\bibinfo {year}
  {2020})}\BibitemShut {NoStop}%
\bibitem [{\citenamefont {Omar}\ \emph {et~al.}(2022)\citenamefont {Omar},
  \citenamefont {Kong}, \citenamefont {Jani}, \citenamefont {Li}, \citenamefont
  {Zhou}, \citenamefont {Lim}, \citenamefont {Prakash}, \citenamefont {Zeng},
  \citenamefont {Hooda}, \citenamefont {Venkatesan}, \citenamefont {Feng},
  \citenamefont {Pennycook}, \citenamefont {Shen},\ and\ \citenamefont
  {Ariando}}]{PhysRevLett.129.187203}%
  \BibitemOpen
  \bibfield  {author} {\bibinfo {author} {\bibfnamefont {G.~J.}\ \bibnamefont
  {Omar}}, \bibinfo {author} {\bibfnamefont {W.~L.}\ \bibnamefont {Kong}},
  \bibinfo {author} {\bibfnamefont {H.}~\bibnamefont {Jani}}, \bibinfo {author}
  {\bibfnamefont {M.~S.}\ \bibnamefont {Li}}, \bibinfo {author} {\bibfnamefont
  {J.}~\bibnamefont {Zhou}}, \bibinfo {author} {\bibfnamefont {Z.~S.}\
  \bibnamefont {Lim}}, \bibinfo {author} {\bibfnamefont {S.}~\bibnamefont
  {Prakash}}, \bibinfo {author} {\bibfnamefont {S.~W.}\ \bibnamefont {Zeng}},
  \bibinfo {author} {\bibfnamefont {S.}~\bibnamefont {Hooda}}, \bibinfo
  {author} {\bibfnamefont {T.}~\bibnamefont {Venkatesan}}, \bibinfo {author}
  {\bibfnamefont {Y.~P.}\ \bibnamefont {Feng}}, \bibinfo {author}
  {\bibfnamefont {S.~J.}\ \bibnamefont {Pennycook}}, \bibinfo {author}
  {\bibfnamefont {L.}~\bibnamefont {Shen}},\ and\ \bibinfo {author}
  {\bibfnamefont {A.}~\bibnamefont {Ariando}},\ }\href
  {https://doi.org/10.1103/PhysRevLett.129.187203} {\bibfield  {journal}
  {\bibinfo  {journal} {Phys. Rev. Lett.}\ }\textbf {\bibinfo {volume} {129}},\
  \bibinfo {pages} {187203} (\bibinfo {year} {2022})}\BibitemShut {NoStop}%
\bibitem [{\citenamefont {Soluyanov}\ and\ \citenamefont
  {Vanderbilt}(2011)}]{PhysRevB.83.235401}%
  \BibitemOpen
  \bibfield  {author} {\bibinfo {author} {\bibfnamefont {A.~A.}\ \bibnamefont
  {Soluyanov}}\ and\ \bibinfo {author} {\bibfnamefont {D.}~\bibnamefont
  {Vanderbilt}},\ }\href {https://doi.org/10.1103/PhysRevB.83.235401}
  {\bibfield  {journal} {\bibinfo  {journal} {Phys. Rev. B}\ }\textbf {\bibinfo
  {volume} {83}},\ \bibinfo {pages} {235401} (\bibinfo {year}
  {2011})}\BibitemShut {NoStop}%
\bibitem [{\citenamefont {Wang}\ \emph {et~al.}(2012)\citenamefont {Wang},
  \citenamefont {Liu}, \citenamefont {Burton}, \citenamefont {Jaswal},\ and\
  \citenamefont {Tsymbal}}]{PhysRevLett.109.247601}%
  \BibitemOpen
  \bibfield  {author} {\bibinfo {author} {\bibfnamefont {Y.}~\bibnamefont
  {Wang}}, \bibinfo {author} {\bibfnamefont {X.}~\bibnamefont {Liu}}, \bibinfo
  {author} {\bibfnamefont {J.~D.}\ \bibnamefont {Burton}}, \bibinfo {author}
  {\bibfnamefont {S.~S.}\ \bibnamefont {Jaswal}},\ and\ \bibinfo {author}
  {\bibfnamefont {E.~Y.}\ \bibnamefont {Tsymbal}},\ }\href
  {https://doi.org/10.1103/PhysRevLett.109.247601} {\bibfield  {journal}
  {\bibinfo  {journal} {Phys. Rev. Lett.}\ }\textbf {\bibinfo {volume} {109}},\
  \bibinfo {pages} {247601} (\bibinfo {year} {2012})}\BibitemShut {NoStop}%
\bibitem [{\citenamefont {Ghosez}\ \emph {et~al.}(1996)\citenamefont {Ghosez},
  \citenamefont {Gonze},\ and\ \citenamefont {Michenaud}}]{ghosez1996}%
  \BibitemOpen
  \bibfield  {author} {\bibinfo {author} {\bibfnamefont {P.}~\bibnamefont
  {Ghosez}}, \bibinfo {author} {\bibfnamefont {X.}~\bibnamefont {Gonze}},\ and\
  \bibinfo {author} {\bibfnamefont {J.-P.}\ \bibnamefont {Michenaud}},\ }\href
  {https://doi.org/10.1209/epl/i1996-00404-8} {\bibfield  {journal} {\bibinfo
  {journal} {EPL}\ }\textbf {\bibinfo {volume} {33}},\ \bibinfo {pages} {713}
  (\bibinfo {year} {1996})}\BibitemShut {NoStop}%
\bibitem [{\citenamefont {Krivovichev}\ \emph {et~al.}(2004)\citenamefont
  {Krivovichev}, \citenamefont {Armbruster},\ and\ \citenamefont
  {Depmeier}}]{krivovichev2004}%
  \BibitemOpen
  \bibfield  {author} {\bibinfo {author} {\bibfnamefont {S.}~\bibnamefont
  {Krivovichev}}, \bibinfo {author} {\bibfnamefont {T.}~\bibnamefont
  {Armbruster}},\ and\ \bibinfo {author} {\bibfnamefont {W.}~\bibnamefont
  {Depmeier}},\ }\href
  {https://doi.org/https://doi.org/10.1016/j.materresbull.2004.05.002}
  {\bibfield  {journal} {\bibinfo  {journal} {Mater. Res. Bull.}\ }\textbf
  {\bibinfo {volume} {39}},\ \bibinfo {pages} {1717} (\bibinfo {year}
  {2004})}\BibitemShut {NoStop}%
\bibitem [{\citenamefont {Gonze}\ and\ \citenamefont
  {et~al.}(2016)}]{gonze2016}%
  \BibitemOpen
  \bibfield  {author} {\bibinfo {author} {\bibfnamefont {X.}~\bibnamefont
  {Gonze}}\ and\ \bibinfo {author} {\bibnamefont {et~al.}},\ }\href
  {https://doi.org/https://doi.org/10.1016/j.cpc.2016.04.003} {\bibfield
  {journal} {\bibinfo  {journal} {Computer Physics Communications}\ }\textbf
  {\bibinfo {volume} {205}},\ \bibinfo {pages} {106 } (\bibinfo {year}
  {2016})}\BibitemShut {NoStop}%
\bibitem [{\citenamefont {Van~Setten}\ \emph {et~al.}(2018)\citenamefont
  {Van~Setten}, \citenamefont {Giantomassi}, \citenamefont {Bousquet},
  \citenamefont {Verstraete}, \citenamefont {Hamann}, \citenamefont {Gonze},\
  and\ \citenamefont {Rignanese}}]{pseudodojo}%
  \BibitemOpen
  \bibfield  {author} {\bibinfo {author} {\bibfnamefont {M.}~\bibnamefont
  {Van~Setten}}, \bibinfo {author} {\bibfnamefont {M.}~\bibnamefont
  {Giantomassi}}, \bibinfo {author} {\bibfnamefont {E.}~\bibnamefont
  {Bousquet}}, \bibinfo {author} {\bibfnamefont {M.~J.}\ \bibnamefont
  {Verstraete}}, \bibinfo {author} {\bibfnamefont {D.~R.}\ \bibnamefont
  {Hamann}}, \bibinfo {author} {\bibfnamefont {X.}~\bibnamefont {Gonze}},\ and\
  \bibinfo {author} {\bibfnamefont {G.-M.}\ \bibnamefont {Rignanese}},\ }\href
  {https://doi.org/https://doi.org/10.1016/j.cpc.2018.01.012} {\bibfield
  {journal} {\bibinfo  {journal} {Comput. Phys. Commun.}\ }\textbf {\bibinfo
  {volume} {226}},\ \bibinfo {pages} {39} (\bibinfo {year} {2018})}\BibitemShut
  {NoStop}%
\bibitem [{\citenamefont {Perdew}\ \emph {et~al.}(1996)\citenamefont {Perdew},
  \citenamefont {Burke},\ and\ \citenamefont {Ernzerhof}}]{Perdew1996}%
  \BibitemOpen
  \bibfield  {author} {\bibinfo {author} {\bibfnamefont {J.~P.}\ \bibnamefont
  {Perdew}}, \bibinfo {author} {\bibfnamefont {K.}~\bibnamefont {Burke}},\ and\
  \bibinfo {author} {\bibfnamefont {M.}~\bibnamefont {Ernzerhof}},\ }\href
  {https://doi.org/10.1103/PhysRevLett.77.3865} {\bibfield  {journal} {\bibinfo
   {journal} {Phys. Rev. Lett.}\ }\textbf {\bibinfo {volume} {77}},\ \bibinfo
  {pages} {3865} (\bibinfo {year} {1996})}\BibitemShut {NoStop}%
\bibitem [{\citenamefont {Perdew}\ \emph {et~al.}(2008)\citenamefont {Perdew},
  \citenamefont {Ruzsinszky}, \citenamefont {Csonka}, \citenamefont {Vydrov},
  \citenamefont {Scuseria}, \citenamefont {Constantin}, \citenamefont {Zhou},\
  and\ \citenamefont {Burke}}]{perdew2008}%
  \BibitemOpen
  \bibfield  {author} {\bibinfo {author} {\bibfnamefont {J.~P.}\ \bibnamefont
  {Perdew}}, \bibinfo {author} {\bibfnamefont {A.}~\bibnamefont {Ruzsinszky}},
  \bibinfo {author} {\bibfnamefont {G.~I.}\ \bibnamefont {Csonka}}, \bibinfo
  {author} {\bibfnamefont {O.~A.}\ \bibnamefont {Vydrov}}, \bibinfo {author}
  {\bibfnamefont {G.~E.}\ \bibnamefont {Scuseria}}, \bibinfo {author}
  {\bibfnamefont {L.~A.}\ \bibnamefont {Constantin}}, \bibinfo {author}
  {\bibfnamefont {X.}~\bibnamefont {Zhou}},\ and\ \bibinfo {author}
  {\bibfnamefont {K.}~\bibnamefont {Burke}},\ }\href
  {https://doi.org/10.1103/PhysRevLett.100.136406} {\bibfield  {journal}
  {\bibinfo  {journal} {Phys. Rev. Lett.}\ }\textbf {\bibinfo {volume} {100}},\
  \bibinfo {pages} {136406} (\bibinfo {year} {2008})}\BibitemShut {NoStop}%
\bibitem [{\citenamefont {Momma}\ and\ \citenamefont
  {Izumi}(2011)}]{momma2011}%
  \BibitemOpen
  \bibfield  {author} {\bibinfo {author} {\bibfnamefont {K.}~\bibnamefont
  {Momma}}\ and\ \bibinfo {author} {\bibfnamefont {F.}~\bibnamefont {Izumi}},\
  }\href {https://doi.org/10.1107/S0021889811038970} {\bibfield  {journal}
  {\bibinfo  {journal} {Journal of Applied Crystallography}\ }\textbf {\bibinfo
  {volume} {44}},\ \bibinfo {pages} {1272} (\bibinfo {year}
  {2011})}\BibitemShut {NoStop}%
\bibitem [{\citenamefont {Herath}\ \emph {et~al.}(2020)\citenamefont {Herath},
  \citenamefont {Tavadze}, \citenamefont {He}, \citenamefont {Bousquet},
  \citenamefont {Singh}, \citenamefont {MuÃ±oz},\ and\ \citenamefont
  {Romero}}]{HERATH2020107080}%
  \BibitemOpen
  \bibfield  {author} {\bibinfo {author} {\bibfnamefont {U.}~\bibnamefont
  {Herath}}, \bibinfo {author} {\bibfnamefont {P.}~\bibnamefont {Tavadze}},
  \bibinfo {author} {\bibfnamefont {X.}~\bibnamefont {He}}, \bibinfo {author}
  {\bibfnamefont {E.}~\bibnamefont {Bousquet}}, \bibinfo {author}
  {\bibfnamefont {S.}~\bibnamefont {Singh}}, \bibinfo {author} {\bibfnamefont
  {F.}~\bibnamefont {MuÃ±oz}},\ and\ \bibinfo {author} {\bibfnamefont
  {A.~H.}\ \bibnamefont {Romero}},\ }\href
  {https://doi.org/https://doi.org/10.1016/j.cpc.2019.107080} {\bibfield
  {journal} {\bibinfo  {journal} {Computer Physics Communications}\ }\textbf
  {\bibinfo {volume} {251}},\ \bibinfo {pages} {107080} (\bibinfo {year}
  {2020})}\BibitemShut {NoStop}%
\bibitem [{\citenamefont {Gresch}\ \emph {et~al.}(2017)\citenamefont {Gresch},
  \citenamefont {Aut\`es}, \citenamefont {Yazyev}, \citenamefont {Troyer},
  \citenamefont {Vanderbilt}, \citenamefont {Bernevig},\ and\ \citenamefont
  {Soluyanov}}]{PhysRevB.95.075146}%
  \BibitemOpen
  \bibfield  {author} {\bibinfo {author} {\bibfnamefont {D.}~\bibnamefont
  {Gresch}}, \bibinfo {author} {\bibfnamefont {G.}~\bibnamefont {Aut\`es}},
  \bibinfo {author} {\bibfnamefont {O.~V.}\ \bibnamefont {Yazyev}}, \bibinfo
  {author} {\bibfnamefont {M.}~\bibnamefont {Troyer}}, \bibinfo {author}
  {\bibfnamefont {D.}~\bibnamefont {Vanderbilt}}, \bibinfo {author}
  {\bibfnamefont {B.~A.}\ \bibnamefont {Bernevig}},\ and\ \bibinfo {author}
  {\bibfnamefont {A.~A.}\ \bibnamefont {Soluyanov}},\ }\href
  {https://doi.org/10.1103/PhysRevB.95.075146} {\bibfield  {journal} {\bibinfo
  {journal} {Phys. Rev. B}\ }\textbf {\bibinfo {volume} {95}},\ \bibinfo
  {pages} {075146} (\bibinfo {year} {2017})}\BibitemShut {NoStop}%
\bibitem [{\citenamefont {Hlinka}\ \emph {et~al.}(2016)\citenamefont {Hlinka},
  \citenamefont {Privratska}, \citenamefont {Ondrejkovic},\ and\ \citenamefont
  {Janovec}}]{PhysRevLett.116.177602}%
  \BibitemOpen
  \bibfield  {author} {\bibinfo {author} {\bibfnamefont {J.}~\bibnamefont
  {Hlinka}}, \bibinfo {author} {\bibfnamefont {J.}~\bibnamefont {Privratska}},
  \bibinfo {author} {\bibfnamefont {P.}~\bibnamefont {Ondrejkovic}},\ and\
  \bibinfo {author} {\bibfnamefont {V.}~\bibnamefont {Janovec}},\ }\href
  {https://doi.org/10.1103/PhysRevLett.116.177602} {\bibfield  {journal}
  {\bibinfo  {journal} {Phys. Rev. Lett.}\ }\textbf {\bibinfo {volume} {116}},\
  \bibinfo {pages} {177602} (\bibinfo {year} {2016})}\BibitemShut {NoStop}%
\bibitem [{\citenamefont {Hayashida}\ \emph {et~al.}(2020)\citenamefont
  {Hayashida}, \citenamefont {Uemura}, \citenamefont {Kimura}, \citenamefont
  {Matsuoka}, \citenamefont {Morikawa}, \citenamefont {Hirose}, \citenamefont
  {Tsuda}, \citenamefont {Hasegawa},\ and\ \citenamefont
  {Kimura}}]{Hayashida2020}%
  \BibitemOpen
  \bibfield  {author} {\bibinfo {author} {\bibfnamefont {T.}~\bibnamefont
  {Hayashida}}, \bibinfo {author} {\bibfnamefont {Y.}~\bibnamefont {Uemura}},
  \bibinfo {author} {\bibfnamefont {K.}~\bibnamefont {Kimura}}, \bibinfo
  {author} {\bibfnamefont {S.}~\bibnamefont {Matsuoka}}, \bibinfo {author}
  {\bibfnamefont {D.}~\bibnamefont {Morikawa}}, \bibinfo {author}
  {\bibfnamefont {S.}~\bibnamefont {Hirose}}, \bibinfo {author} {\bibfnamefont
  {K.}~\bibnamefont {Tsuda}}, \bibinfo {author} {\bibfnamefont
  {T.}~\bibnamefont {Hasegawa}},\ and\ \bibinfo {author} {\bibfnamefont
  {T.}~\bibnamefont {Kimura}},\ }\href
  {https://doi.org/10.1038/s41467-020-18408-6} {\bibfield  {journal} {\bibinfo
  {journal} {Nat. Commun.}\ }\textbf {\bibinfo {volume} {11}},\ \bibinfo
  {pages} {4582} (\bibinfo {year} {2020})}\BibitemShut {NoStop}%
\bibitem [{\citenamefont {Iwata}(1977)}]{doi:10.1143/JPSJ.43.961}%
  \BibitemOpen
  \bibfield  {author} {\bibinfo {author} {\bibfnamefont {Y.}~\bibnamefont
  {Iwata}},\ }\href {https://doi.org/10.1143/JPSJ.43.961} {\bibfield  {journal}
  {\bibinfo  {journal} {Journal of the Physical Society of Japan}\ }\textbf
  {\bibinfo {volume} {43}},\ \bibinfo {pages} {961} (\bibinfo {year}
  {1977})}\BibitemShut {NoStop}%
\bibitem [{\citenamefont {Newnham}\ \emph {et~al.}(1973)\citenamefont
  {Newnham}, \citenamefont {Wolfe},\ and\ \citenamefont
  {Darlington}}]{newnham1973}%
  \BibitemOpen
  \bibfield  {author} {\bibinfo {author} {\bibfnamefont {R.}~\bibnamefont
  {Newnham}}, \bibinfo {author} {\bibfnamefont {R.}~\bibnamefont {Wolfe}},\
  and\ \bibinfo {author} {\bibfnamefont {C.}~\bibnamefont {Darlington}},\
  }\href {https://doi.org/10.1016/0022-4596(73)90226-0} {\bibfield  {journal}
  {\bibinfo  {journal} {J. Solid State Chem.}\ }\textbf {\bibinfo {volume}
  {6}},\ \bibinfo {pages} {378 } (\bibinfo {year} {1973})}\BibitemShut
  {NoStop}%
\bibitem [{\citenamefont {Gonze}\ and\ \citenamefont {Lee}(1997)}]{gonze1997}%
  \BibitemOpen
  \bibfield  {author} {\bibinfo {author} {\bibfnamefont {X.}~\bibnamefont
  {Gonze}}\ and\ \bibinfo {author} {\bibfnamefont {C.}~\bibnamefont {Lee}},\
  }\href {https://doi.org/10.1103/PhysRevB.55.10355} {\bibfield  {journal}
  {\bibinfo  {journal} {Phys. Rev. B}\ }\textbf {\bibinfo {volume} {55}},\
  \bibinfo {pages} {10355} (\bibinfo {year} {1997})}\BibitemShut {NoStop}%
\bibitem [{\citenamefont {Bousquet}\ and\ \citenamefont
  {Ghosez}(2006)}]{bousquet2006}%
  \BibitemOpen
  \bibfield  {author} {\bibinfo {author} {\bibfnamefont {E.}~\bibnamefont
  {Bousquet}}\ and\ \bibinfo {author} {\bibfnamefont {P.}~\bibnamefont
  {Ghosez}},\ }\href {https://doi.org/10.1103/PhysRevB.74.180101} {\bibfield
  {journal} {\bibinfo  {journal} {Phys. Rev. B}\ }\textbf {\bibinfo {volume}
  {74}},\ \bibinfo {pages} {180101} (\bibinfo {year} {2006})}\BibitemShut
  {NoStop}%
\bibitem [{\citenamefont {Viennois}\ \emph {et~al.}(2018)\citenamefont
  {Viennois}, \citenamefont {Kityk}, \citenamefont {Majchrowski}, \citenamefont
  {Zmija}, \citenamefont {Mierczyk},\ and\ \citenamefont
  {Papet}}]{VIENNOIS2018461}%
  \BibitemOpen
  \bibfield  {author} {\bibinfo {author} {\bibfnamefont {R.}~\bibnamefont
  {Viennois}}, \bibinfo {author} {\bibfnamefont {I.}~\bibnamefont {Kityk}},
  \bibinfo {author} {\bibfnamefont {A.}~\bibnamefont {Majchrowski}}, \bibinfo
  {author} {\bibfnamefont {J.}~\bibnamefont {Zmija}}, \bibinfo {author}
  {\bibfnamefont {Z.}~\bibnamefont {Mierczyk}},\ and\ \bibinfo {author}
  {\bibfnamefont {P.}~\bibnamefont {Papet}},\ }\href
  {https://doi.org/https://doi.org/10.1016/j.matchemphys.2018.04.025}
  {\bibfield  {journal} {\bibinfo  {journal} {Mater. Chem. Phys.}\ }\textbf
  {\bibinfo {volume} {213}},\ \bibinfo {pages} {461} (\bibinfo {year}
  {2018})}\BibitemShut {NoStop}%
\bibitem [{\citenamefont {Hussain}\ \emph {et~al.}(2021)\citenamefont
  {Hussain}, \citenamefont {Rashid}, \citenamefont {Saeed},\ and\ \citenamefont
  {Bhatti}}]{hussain2021}%
  \BibitemOpen
  \bibfield  {author} {\bibinfo {author} {\bibfnamefont {M.}~\bibnamefont
  {Hussain}}, \bibinfo {author} {\bibfnamefont {M.}~\bibnamefont {Rashid}},
  \bibinfo {author} {\bibfnamefont {F.}~\bibnamefont {Saeed}},\ and\ \bibinfo
  {author} {\bibfnamefont {A.~S.}\ \bibnamefont {Bhatti}},\ }\href
  {https://doi.org/10.1007/s10853-020-05298-8} {\bibfield  {journal} {\bibinfo
  {journal} {J. Mater. Sci.}\ }\textbf {\bibinfo {volume} {56}},\ \bibinfo
  {pages} {528} (\bibinfo {year} {2021})}\BibitemShut {NoStop}%
\bibitem [{\citenamefont {Bahramy}\ \emph {et~al.}(2011)\citenamefont
  {Bahramy}, \citenamefont {Arita},\ and\ \citenamefont
  {Nagaosa}}]{PhysRevB.84.041202}%
  \BibitemOpen
  \bibfield  {author} {\bibinfo {author} {\bibfnamefont {M.~S.}\ \bibnamefont
  {Bahramy}}, \bibinfo {author} {\bibfnamefont {R.}~\bibnamefont {Arita}},\
  and\ \bibinfo {author} {\bibfnamefont {N.}~\bibnamefont {Nagaosa}},\ }\href
  {https://doi.org/10.1103/PhysRevB.84.041202} {\bibfield  {journal} {\bibinfo
  {journal} {Phys. Rev. B}\ }\textbf {\bibinfo {volume} {84}},\ \bibinfo
  {pages} {041202} (\bibinfo {year} {2011})}\BibitemShut {NoStop}%
\bibitem [{\citenamefont {Elliott}(1954)}]{PhysRev.96.266}%
  \BibitemOpen
  \bibfield  {author} {\bibinfo {author} {\bibfnamefont {R.~J.}\ \bibnamefont
  {Elliott}},\ }\href {https://doi.org/10.1103/PhysRev.96.266} {\bibfield
  {journal} {\bibinfo  {journal} {Phys. Rev.}\ }\textbf {\bibinfo {volume}
  {96}},\ \bibinfo {pages} {266} (\bibinfo {year} {1954})}\BibitemShut
  {NoStop}%
\bibitem [{\citenamefont {Yafet}(1963)}]{YAFET19631}%
  \BibitemOpen
  \bibfield  {author} {\bibinfo {author} {\bibfnamefont {Y.}~\bibnamefont
  {Yafet}}\ }(\bibinfo  {publisher} {Academic Press},\ \bibinfo {year} {1963})\
  pp.\ \bibinfo {pages} {1--98}\BibitemShut {NoStop}%
\bibitem [{\citenamefont {{D'Yakonov}}\ and\ \citenamefont
  {{Perel'}}(1971)}]{1971JETP...33.1053D}%
  \BibitemOpen
  \bibfield  {author} {\bibinfo {author} {\bibfnamefont {M.~I.}\ \bibnamefont
  {{D'Yakonov}}}\ and\ \bibinfo {author} {\bibfnamefont {V.~I.}\ \bibnamefont
  {{Perel'}}},\ }\href@noop {} {\bibfield  {journal} {\bibinfo  {journal}
  {Soviet Journal of Experimental and Theoretical Physics}\ }\textbf {\bibinfo
  {volume} {33}},\ \bibinfo {pages} {1053} (\bibinfo {year}
  {1971})}\BibitemShut {NoStop}%
\bibitem [{\citenamefont {Bernevig}\ \emph {et~al.}(2006)\citenamefont
  {Bernevig}, \citenamefont {Orenstein},\ and\ \citenamefont
  {Zhang}}]{PhysRevLett.97.236601}%
  \BibitemOpen
  \bibfield  {author} {\bibinfo {author} {\bibfnamefont {B.~A.}\ \bibnamefont
  {Bernevig}}, \bibinfo {author} {\bibfnamefont {J.}~\bibnamefont
  {Orenstein}},\ and\ \bibinfo {author} {\bibfnamefont {S.-C.}\ \bibnamefont
  {Zhang}},\ }\href {https://doi.org/10.1103/PhysRevLett.97.236601} {\bibfield
  {journal} {\bibinfo  {journal} {Phys. Rev. Lett.}\ }\textbf {\bibinfo
  {volume} {97}},\ \bibinfo {pages} {236601} (\bibinfo {year}
  {2006})}\BibitemShut {NoStop}%
\bibitem [{\citenamefont {Bruneval}\ \emph {et~al.}(2015)\citenamefont
  {Bruneval}, \citenamefont {Varvenne}, \citenamefont {Crocombette},\ and\
  \citenamefont {Clouet}}]{bruneval2015}%
  \BibitemOpen
  \bibfield  {author} {\bibinfo {author} {\bibfnamefont {F.}~\bibnamefont
  {Bruneval}}, \bibinfo {author} {\bibfnamefont {C.}~\bibnamefont {Varvenne}},
  \bibinfo {author} {\bibfnamefont {J.-P.}\ \bibnamefont {Crocombette}},\ and\
  \bibinfo {author} {\bibfnamefont {E.}~\bibnamefont {Clouet}},\ }\href
  {https://doi.org/10.1103/PhysRevB.91.024107} {\bibfield  {journal} {\bibinfo
  {journal} {Phys. Rev. B}\ }\textbf {\bibinfo {volume} {91}},\ \bibinfo
  {pages} {024107} (\bibinfo {year} {2015})}\BibitemShut {NoStop}%
\end{thebibliography}%

\newpage

\section{Supplementary Material}
\beginsupplement

\subsection{Technical Details}

The density functional theory calculations have been performed with the ABINIT code (v9.6.2)~\cite{gonze2016,gonze2020} and through norm-conserving pseudo-potentials from the PseudoDojo project~\cite{pseudodojo} (v0.4).
The Perdew-Burke-Ernzerhof revised for solids (PBEsol) functional~\cite{Perdew1996,perdew2008} was employed within the generalised gradient approximation (GGA). The lattice parameters of PGO have also been estimated via PBE and LDA functionals.
To sample the Brillouin zone, a 3x3x3 k-point grid and a plane-wave cutoff of 40 Ha were used.
The projected density and partial charge figures were obtained with the VASP (v5.4) code~\cite{Kresse1999} and the VESTA software~\cite{momma2011}. 
The spin texture has been extracted with the help of  PyProcar (v5.6.6)~\cite{HERATH2020107080}.
The Chern number associated with the lowest unoccupied conduction band states has been evaluated with the help of the Z2pack code~\cite{PhysRevB.95.075146,PhysRevB.83.235401} and analytical modelling.

\subsection{Structural information}

The paraelectric phase of PGO, which Wyckoff positions are listed in tab.~\ref{tab:PE-WPs}, is also described in the main section. As for the ferroelectric phase, among its 23 WPs the 3d site is occupied by all the oxygen and germanium atoms, while the lead sites appear in both 3d, 1c and 1b positions.

\begin{table}[htbp!]
\begin{center}
\begin{tabular}{ccccc}
\hline
\hline
 Atom  & site sym. & $x$ & $y$ & $z$ \rule{0pt}{8pt}\\
\hline
Pb$_1$ & $3k$ & 0.2527 & -0.0033 &  1/2 \rule{0pt}{8pt}\\
Pb$_2$ & $6l$ & 0.0027 & -0.2625 & 0.1812 \\
Pb$_3$ & $1e$ & 2/3 & 1/3 & 0.0073 \\
Pb$_4$ & $2i$ & 2/3 & 1/3 & 0.3383 \\
Pb$_5$ & $1c$ & 1/3 & 2/3 & 0.0039 \\
Pb$_6$ & $2h$ & 1/3 & 2/3 & 0.3327 \\
Ge$_1$ & $6l$ & 0.3863 & 0.0053 & 0.1621 \\
Ge$_2$ & $3k$ & 0.0038 & 0.6122 & 0.4957 \\
O$_1$  & $6l$ & 0.0855 & 0.6862 & 0.6419 \\
O$_2$  & $3k$ & 0.4855 & 0.2889 & 0.5006 \\
O$_3$  & $3k$ & 0.0755 & 0.4974 & 0.4447  \\
O$_4$  & $6l$ & 0.1917 & 0.4756 & 0.1705  \\
O$_5$  & $3j$ & 0.3536 & 0.0677 & 0.0111  \\
O$_6$  & $6l$ & 0.3122 & 0.0821 & 0.2687 \\
\hline
\end{tabular}
\caption{Relaxed atomic positions of the P$\Bar{6}$ phase of PGO.}
\label{tab:PE-WPs}
\end{center}
\end{table}

It can also be noticed that both $P\Bar{6}$ and $P3$ space groups present an axial order~\cite{PhysRevLett.116.177602,Hayashida2020} and a non-zero piezoelectric tensor, which is less common for the PE phase of ferroelectric crystals.
In tab.~\ref{tab-acell} we report the lattice parameters, calculated via LDA, PBE and PBEsol functionals, along with their experimental values. Thus we can compare our numerical results with measurements. If we consider the PE phase, the error over the $a$ parameter $\Delta{a}_\text{expt.}$ takes the following values: -2.14 \% (LDA), 1.69 \% (PBE, no SOC), 2.06 \% (PBE with SOC), -0.45 \% (PBEsol, no SOC), -0.38 \% (PBEsol, with SOC). Conversely, the errors over the $c$ parameter $\Delta{c}_\text{expt.}$ are: -1.51 \% (LDA, no SOC), 1.82 \% (PBE, no SOC), 1.75 \% (PBE with SOC), -0.13 \% (PBEsol, no SOC) and -0.25 \% (PBEsol, with SOC).  Repeating the calculations of $\Delta{a}_\text{expt.}$ and $\Delta{a}_\text{expt.}$ in the low symmetry phase reveals: -1.16 \% (LDA), 2.61 \% (PBE, no SOC), 2.82 \% (PBE with SOC), 0.54 \% (PBEsol, no SOC), 0.66 \% (PBEsol, with SOC) in the $a$ case and  -0.86 \% (LDA, no SOC), 3.01 \% (PBE, with and without SOC), 0.71 \% (PBEsol, no SOC) and 0.61 \% (PBEsol, with SOC) when $c$ is considered.
Clearly, the PBEsol functional gives a better estimation of both lattice parameters, which is the reason why it has been adopted throughout our calculations. We can estimate the PBEsol strain associated with the phase transition as $(a_\text{FE}-a_\text{PE})/a_\text{PE}$ = 0.3 \% (no SOC)-0.35 \% (with SOC) and $(c_\text{FE}-c_\text{PE})/c_\text{PE}$ = 0.16 \% (no SOC) - 0.19 \%. If we define an average strain S$_\text{av}$ $\equiv$ $\frac{1}{3}\Big[2\frac{\delta{a}}{a_\text{PE}} + \frac{\delta{c}}{c_\text{PE}}\Big]$, we can see that S$_\text{av}$ = 0.2 \% (no SOC) - 0.3 \% (with SOC).

\begin{table}[htbp!]
\begin{center}
\begin{tabular}{lcccc}
\hline
\hline
Exc && $a$ & $c$ & Ref. \\
\hline
\multicolumn{5}{c}{PE phase} \\
LDA (wo SOC)   && 10.040 & 10.534 & \\
PBE (wo SOC)   && 10.433 & 10.891 & \\
PBE (w SOC)   && 10.471 & 10.883 & \\
PBEsol (wo SOC) && 10.214 & 10.682 & \\
PBEsol (w SOC) && 10.221 & 10.669 & \\
Expt.    && 10.260 &  10.696 &  \cite{doi:10.1143/JPSJ.43.961} (200 K) \\
\hline
\multicolumn{5}{c}{FE phase} \\
LDA (wo SOC)    && 10.072 & 10.530 & \\
PBE (wo SOC)    && 10.456 & 10.944 & \\
PBE (w SOC)     && 10.477 & 10.944 & \\
PBEsol (wo SOC) && 10.245 & 10.699 & \\
PBEsol (w SOC) && 10.257 & 10.689 & \\
Expt.            && 10.251 &  10.685 &  \cite{iwasaki1971} (RT) \\
Expt.     && 10.190 &  10.624 &  \cite{newnham1973} (RT) \\

\hline
\end{tabular}
\label{tab-acell}
\caption{Cell parameters (\AA) of PE and FE phase of PGO as obtained through different exchange correlation functionals (Exc) and comparison with experiments (RT means room temperature measurements).}
\end{center}
\end{table}

Finally, our calculations reveal that the lattice relaxation of the ferroelectric ground state amounts to a gain of only 8 meV, which means it cannot be considered the main source of the SOC-induced renormalisation of $\Delta$E reported in the main text.

\subsection{Paraelectric soft mode frequency calculation}

The phonon frequencies at $\Gamma$ in the P$\Bar{6}$ phase are computed via density functional perturbation theory (DFPT)~\cite{gonze1997}. Our result indicate that both the TO and LO frequencies are soft (34.8i cm$^{-1}$ and 28i cm$^{-1}$ respectively). 
Following refs.~\cite{ghosez1996, bousquet2006}, the $\Gamma$-point dynamical matrix is divided into a short (SR)- and a long (LR)-range contribution which generate the $\omega_{0;SR}^{2}$ and $\omega_{0;LR}^{2}$ squared frequencies respectively. We find $\omega_{0;SR}^{2}$ = -2235.2 cm$^{-1}$ and $\omega_{0;LR}^{2}$ = 1022.5 cm$^{-1}$ 
which clearly indicates that the polar instability has a short-range nature.


\subsection{Electronic properties of the paraelectric phase}


\begin{figure}
 \centering
 \includegraphics[width=12cm,keepaspectratio=true]{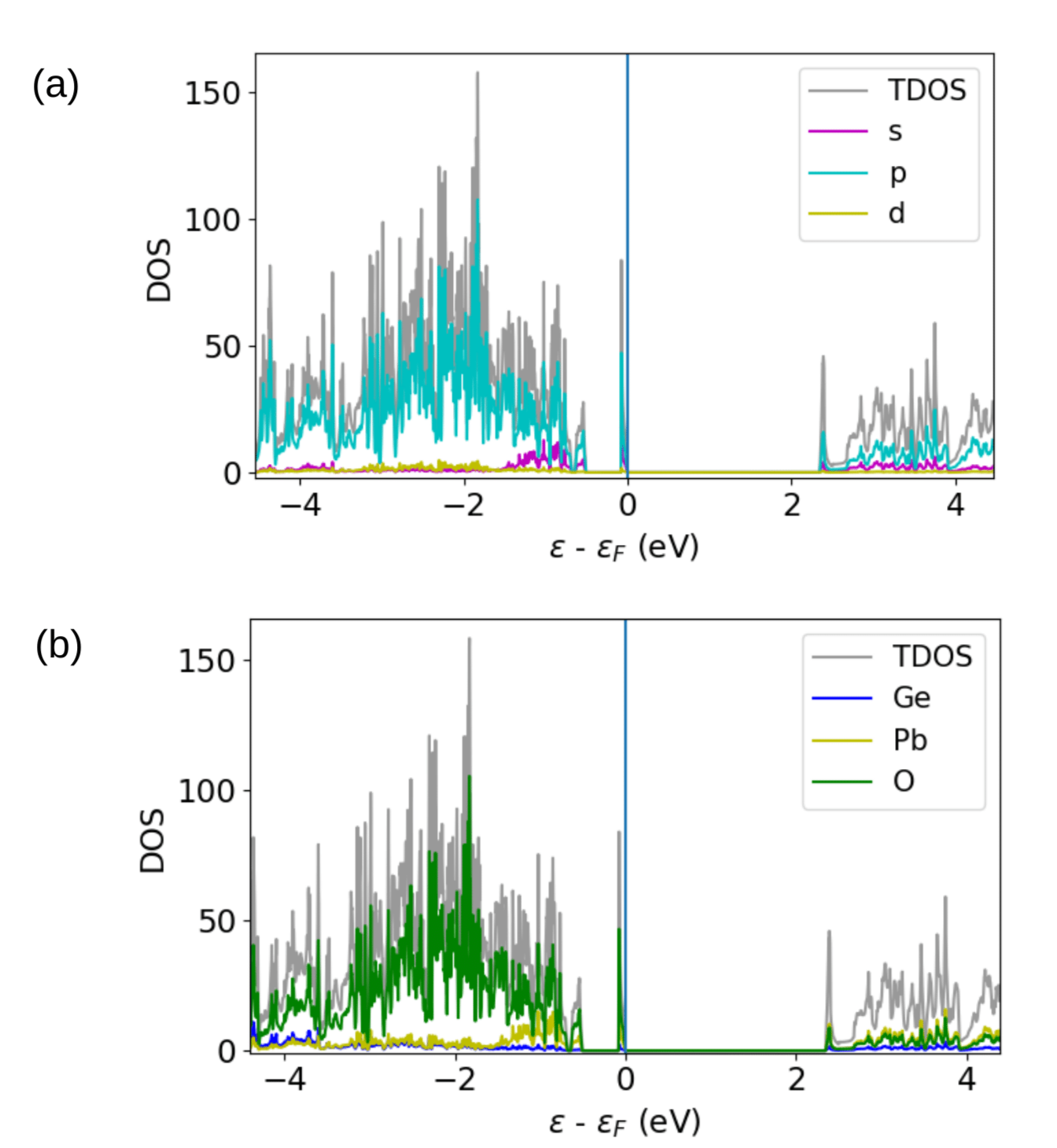}
 \caption{Orbital (a) and atom (b) projected density of states of the P$\Bar{6}$ phase in absence of SOC.}
 \label{fig:DOS_PE_merge}
\end{figure}

\begin{figure}[ht]
     \includegraphics[width=0.8\textwidth]{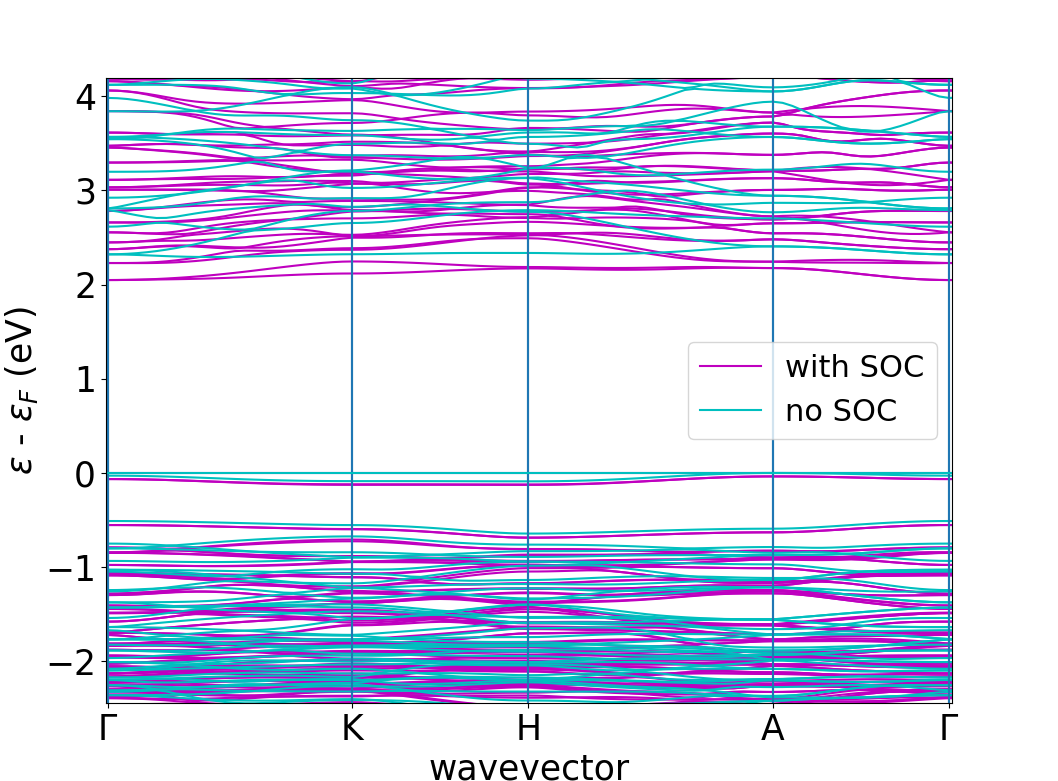}
     \caption{P$\Bar{6}$ band structure zoomed around the Fermi level.}
     \label{Fig:PE_fullband}
     \end{figure}

In fig.~\ref{fig:DOS_PE_merge} we report our calculated density of states in the PE reference phase, both orbital and atom resolved. Much like the low symmetry phase in the main text, the valence bands are dominated by the contribution of O-2p levels, with a small contributions from Pb-6s at the Fermi level and Pb-6p. The top-VB peak has been suggested to play a relevant role in determining the optical properties of PGO~\cite{VIENNOIS2018461}, including a nonlinear optical effect upon chromium doping.
The CB has instead a much larger weight coming from lead atoms.
As in the P3 phase, this results in a bigger SOC-induced splitting of the conduction levels as can be observed in fig.~\ref{Fig:PE_fullband}. The (p$_{x}$,p$_{y}$) bottom CB states (without SOC) belonging to the E' single representation of C$_{3h}$ split according to the $\Bar{E}_{1}$ and $\Bar{E}_{2}$ double IR, which correspond to the crystal field split J = 3/2 levels. The $\Gamma_{1}$ single IR is obviously mapped to a singlet.

\begin{figure}
 \centering
 \includegraphics[width=9cm,keepaspectratio=true]{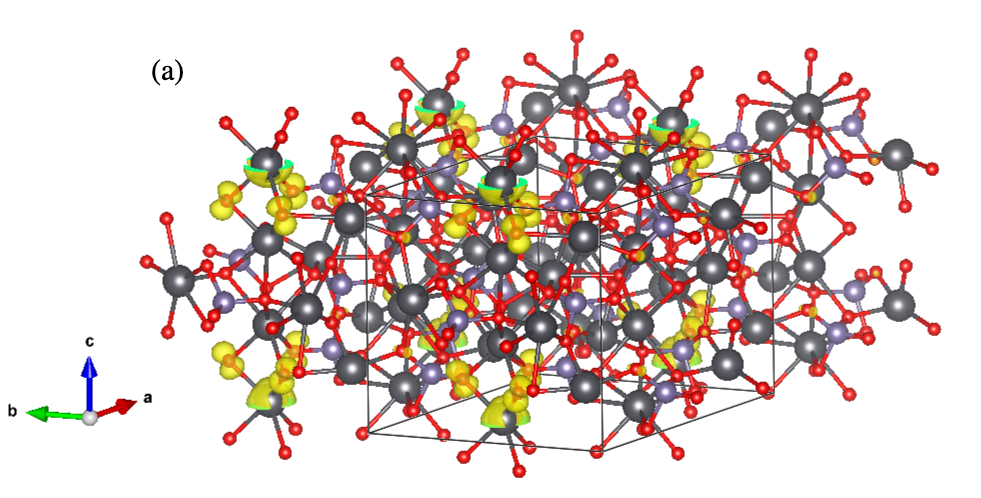}
 \includegraphics[width=9cm,keepaspectratio=true]{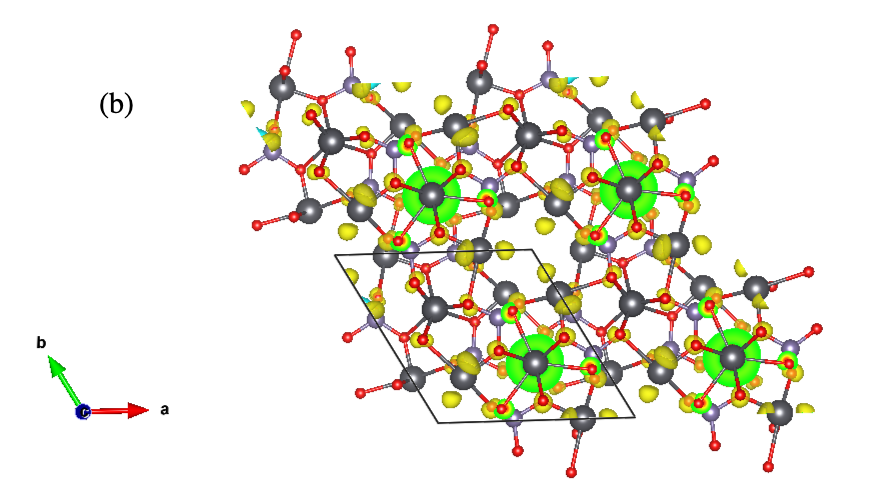}
 \caption{Spatial charge distribution of the (a) top VB and (b) bottom CB states in the paraelectric phase.}
 \label{fig:DOS_PARCHG_PE_modified}
\end{figure}

In fig.~\ref{fig:DOS_PARCHG_PE_modified}(a) and (b) we highlight the charge distribution of the top valence and bottom conduction states respectively. The top-VB has the largest contribution coming from the O-6l and Pb-1c Wyckoff positions and shows the sign of $sp$ hybridisation.
The involved oxygen atoms also bridge Pb-1c, Pb-2h and Ge-6l WPs. These states play a paramount role in stabilising the ferroelectric phase: indeed, depopulating them (holes concentration $>$ 0.655 holes f.u.) results in the cancellation of the phase transition as can be noticed in the main text.
On the other hand, the bottom-CB density is centered around O-(3k,6l), Pb-1e and in the vacuum region which neighbours the Pb-6l WPs.

\subsection{Symmetry analysis of the $\Gamma$-point states}

If we look at the P$\Bar{6}$ phase (point group C$_{\text{3h}}$) with SOC switched off, the lowest two bands belong respectively to the E' = ${\Gamma}_{3}$ 
$\oplus$ ${\Gamma}_{5}$ ($\sim$ p$_{x}$, p$_{y}$) and $\Gamma_{1}$ representations of the single group with $\epsilon[{E'}]<\epsilon[{{\Gamma_{1}}}]$. With spin-orbit, these bands split according to the double group IR and following 
D$_{3/2}$ = E' $\otimes$ D$_{1/2}$ = $\Bar{E}_{1}$ $\oplus$ $\Bar{E}_{2}$ and
D$_{1/2}$ = D$_{1/2}$ $\otimes$ $\Bar{\Gamma}_{1}$ = E$_{3}$, where D is the SO(3) representation of the spin and $\epsilon[{\Bar{E}_{2}}]<\epsilon[{\Bar{E}_{1}}]<\epsilon[{D_{1/2}}]$. 
If instead we analyse the lowest $\Gamma$-CB states of the P3 structure (point group C$_{3}$) without SOC, we see that they belong to the E = ${\Gamma}_{2}$ $\oplus$ ${\Gamma}_{3}$ and $\Gamma_{1}$ single IRs of C$_{3}$, with $\epsilon[{E}]<\epsilon[{\Gamma_{1}}]$. 
Switching on the spin-orbit interaction, these single group representations split according to D$_{3/2}$ = E $\otimes$ D$_{1/2}$ = $\Bar{\Gamma}_{4}$ $\oplus$ $\Bar{\Gamma}_{5,6}$ and D$_{1/2}$ = D$_{1/2}$ $\otimes$ ${\Gamma_{1}}$ = $\Bar{\Gamma}_{5,6}$ with  $\epsilon_{\Bar{\Gamma}_{5,6}}<\epsilon_{\Bar{\Gamma}_{4}}<\epsilon[{D_{1/2}}]$ ( $\Bar{\Gamma}_{5,6}\equiv\Bar{\Gamma}_{5}\oplus\Bar{\Gamma}_{6}$). Therefore, the (crystal field split) J = 3/2 states are at the bottom of the conduction band in both the high and low symmetry phases of PGO.

We point out that a large $p$-dominated conduction bands splitting is also observed in halide perovskites like CsPbBr$_3$~\cite{hussain2021}, although in that case it is the singlet state that appears at the bottom of the CB as a result of a band inversion induced by the SOC. 
To test whether a similar effect occurs in PGO as well, we have performed a symmetry analysis of the bands with an artificially reduced spin-orbit strength. We have adopted, arbitrarily, a scaling factor of 0.05 for the SOC. With respect to the full SOC case, we find the ordering of the bands highlighted in fig. 2 in the main text to be preserved, meaning that there is no singlet-triplet inversion as in CsPbBr$_3$.

The P$\Bar{6}$ axial vectors (no SOC) can be decomposed according to $A'$ $\oplus$ $E"$, meaning that when using the single group IRs (CBM) as the unperturbed basis, the in-plane elements of the average angular momentum ($\braket{\mathbf{L}}$) are not zero since they contain the invariant IR. Likewise, the P3 axial vectors (again without SOC) have components over the two irreducible representations of C$_{3}$, and it can be easily shown that both $\braket{\mathbf{L_{xy}}}$ and $\braket{{L_{z}}}$ are non-zero when averaged over the CBM states.
Since the average orbital angular momentum is not quenched, the SOC ($\sim \braket{\mathbf{L}}\cdot\mathbf{S}$) acts as a first order effect on the energy landscape rather than a second correction as it would happen with quenched $\mathbf{L}$-compounds like, for instance, BiTeI~\cite{PhysRevB.84.041202}.

\subsection{$\mathbf{k}\cdot\mathbf{p}$ model and P3 spin texture}

\begin{figure}[ht]
     \includegraphics[width=0.7\textwidth]{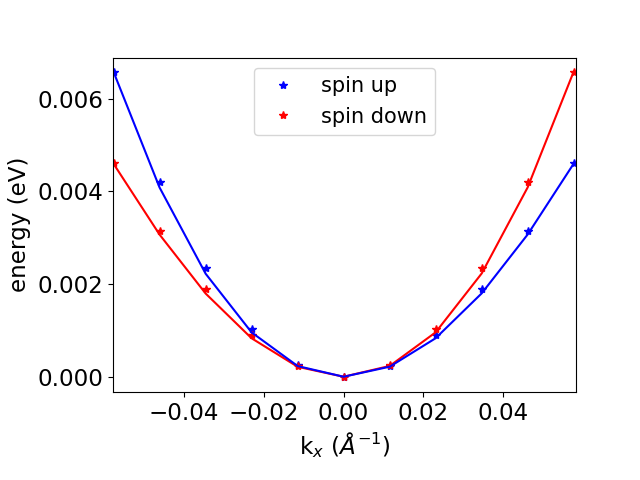}
     \caption{Bottom conduction spin bands in the P$\Bar{6}$ phase. Star symbols are density functional theory calculations while the continuous line is the fitted $\mathbf{k}\cdot\mathbf{p}$ model.}
     \label{Fig:PE_band}
     \end{figure}

In this section we report the $\mathbf{k}\cdot{\mathbf{p}}$ parameters with  reference to the bottom conduction band states in presence of spin-orbit interaction (active for all Wyckoff sites). The pristine phase is assumed.
The high symmetry phase is cubic in $\mathbf{k}$ as found in ref.~\cite{PhysRevLett.125.216405}. The Hamiltonian (k$_{z}$ = 0) is given by the following equation, at leading order:


\begin{equation}\label{PE_k_dot_p}
H_{P\Bar{6}}(k_{x},k_{y},k_{z}=0)  =  \frac{\hbar^{2}}{2m^{*}_{{\text{PE}},(xy)}}(k_{x}^{2}+k_{y}^{2}) + \frac{\hbar^{2}}{2m^{*}_{{\text{PE}},(z)}}k_z^2 + d_\text{PE}(k_x,k_y)\sigma_{z} 
\end{equation}

with

\begin{equation}
    d_\text{PE}(k_x,k_y) \equiv Ck_{x}(k_{x}^{2} - 3k_{y}^{2}) + Dk_{y}(3k_{x}^{2} - k_{y}^{2})
\end{equation}

where m$^{*}_{\text{PE(xy)}}$, m$^{*}_{\text{PE(z)}}$, C and D are parameters of the model. We set $k_z$ = 0. With reference to the states at the bottom of the conduction bands and in presence of SOC, these parameters (fitted from our first principles calculations) are as follow: m$^{*}_{\text{PE(xy)}}$ = 6 m$_{e}$, C = 5.18 eV$\cdot$Angstrom$^{3}$ and D = 3.36 eV$\cdot$Angstrom$^{3}$. In particular, the values of C and D are very close to the estimates of ref.~\cite{PhysRevLett.125.216405}. The plot of the corresponding band structure is given in fig.~\ref{Fig:PE_band}.

A persistent spin texture can be expected since the eigenstates of eq.~\ref{PE_k_dot_p} are momentum independent. It is quite clear that the SOC results in a persistent spin texture (PST) with U(1) unitary symmetry (generated by $\sigma_{z}$), which guarantees a certain robustness upon $\mathbf{k}$ scattering events~\cite{PhysRev.96.266,YAFET19631,1971JETP...33.1053D}.
We further stress that the paraelectric PST is due to the presence of the mirror m$_{001}$ symmetry along with the three-fold proper rotations and roto-reflections characterising the C$_{3h}$ group, which factor out linear SOC-contributions in $\mathbf{k}$. 
Thus, PGO does not have the SU(2) protection encountered when Rashba and Dresselhaus effects compensate or in the (110) Dresselhaus model~\cite{PhysRevLett.97.236601}, which means that two-body operators can still cause spin-momentum scattering. Therefore, we can expect limitations on the lifetime of the spin states even in absence of higher order terms of the $\mathbf{k}\cdot\mathbf{p}$ expansion.
Finally, we remark that the absence of non-symmorphic operations ensures a different PST mechanism with respect to the Kramers protection resulting from the coupling between time reversal and improper rotations~\cite{Tao2018}.


The P3 phase Hamiltonian, already introduced in the main section, reads:


\begin{equation}\label{FE_k_dot_p_SI}
    H_{\text{P3}}(k_{x},k_{y},k_{z}) = \frac{\hbar^{2}}{2m^{*}_{\text{FE;(xy))}}}(k_{x}^{2}+k_{y}^{2}) +\frac{\hbar^{2}}{2m^{*}_{\text{FE;z}}}k_z^2 + \mathbf{d}_\text{FE}(k_x,k_y,k_z)\cdot\mathbf{\sigma},
\end{equation}

with 

  
\begin{align}
    \mathbf{d}(k_x,k_y) &\equiv \begin{bmatrix}
           \lambda_{R}k_{y} + \lambda_{w}k_{x} \\
           -\lambda_{R}k_{x} + \lambda_{w}k_{y} \\
           d_\text{PE}(k_x,k_y) + \lambda_{W_z}k_z
         \end{bmatrix}.
  \end{align}

\begin{figure}[ht]
     \includegraphics[width=0.7\textwidth]{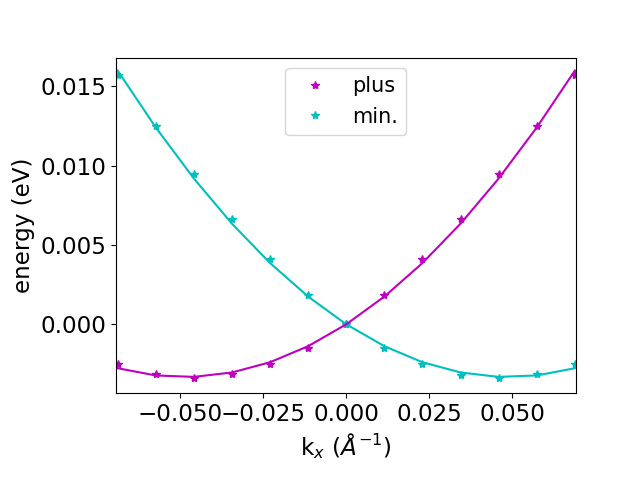}
     \caption{Bottom conduction spin bands, P3 phase. Again, symbols represent DFT results while the continuous line is the interpolated model.}
     \label{Fig:FE_band}
     \end{figure}

Clearly, the $\lambda_{R}$ parameter is the coupling to the Rashba type of spin-orbit, while the $\{\lambda_{W_i}\}$ label a Weyl crossing which is persistent upon application of a Zeeman term. The spectrum is now radically different in comparison with the PE reference. Breaking the mirror symmetry generates linear terms proportional to the in-plane components of the spin polarisation. 
These terms obviously destroy the PST of the P$\Bar{6}$ phase since the states acquire a momentum dependency. 
The electronic dispersion associated with eq.~\ref{FE_k_dot_p_SI} reads  $\epsilon(k)_{\pm} = \pm \sqrt{\alpha^2{(k_x^2+k_y^2) + (\lambda_{W_z}k_z)^2}}$, with $\alpha^2 \equiv \lambda_{R}^2 + \lambda_{w}^2$. In analogy with ref.~\cite{Djani2019} we define $\alpha$ = $2$E$_{R}$/k$_{R}$. From density functional theory simulations we find m$_{\text{FE(xy)}}$ = 7.66 m$_{e}$ and $|\alpha| \equiv \sqrt{\lambda_{R}^2+\lambda_{w}^2}$ = 0.15 eV$\cdot$Angstrom. A plot of the band structure is given if fig.~\ref{Fig:FE_band}.

\begin{figure}
     \includegraphics[width=0.8\textwidth]{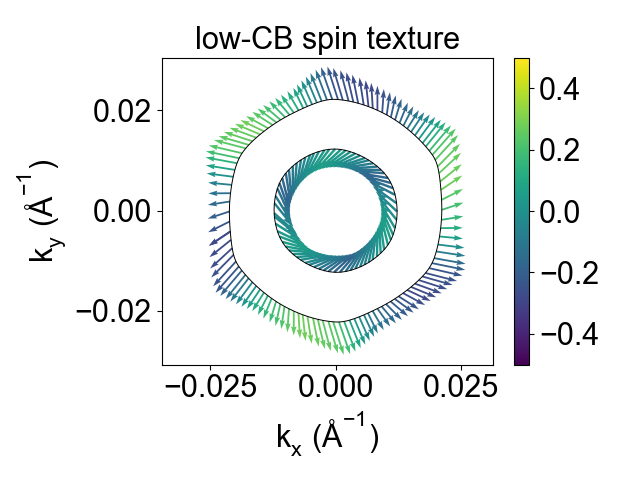}
     \caption{Chiral spin texture (bottom conduction bands) of the ferroelectric phase of PGO.}
     \label{fig:FE_dn_texture}
     \end{figure}

In this case the eigenfunctions have the form $\ket{\pm} = \frac{1}{\sqrt{2}}(\pm e^{i\phi},1)$, with $\phi = \phi(k_x,k_y)\equiv \phi_{\mathbf{k}}$. If we average the spin over these states, we can show that $\braket{\mathbf{S}}_{\pm}$ = $\pm\frac{\hbar}{2}(\cos(\phi_{\mathbf{k}}),\sin(\phi_{\mathbf{k}}),0)$. Unlike the paraelectric case, fitting the energy bands as a function of $\mathbf{k}$ allows for the resolution of $\alpha$ but not of $\lambda_{R}$ and $\lambda_{w}$ separately (up to a sign factor) due to the isotropic planar dispersion. Therefore, we are unable to analytically guess the spin texture from the knowledge of the bandstructure alone and we need to evaluate $\phi(k_x,k_y)$ directly from DFT. The first principles computed spin texture from which we can extract $\tan[\phi(k_x,k_{y})]$
is reproduced in fig.~\ref{fig:FE_dn_texture}. We can notice the onset of a hexagonal warping, produced by symmetry allowed cubic terms not included in eq.\ref{PE_k_dot_p}.
Since $\frac{\braket{{S}}_{y}}{\braket{{S}}_{x}}$ $\rightarrow$ $\frac{\lambda_{w} - \lambda_{R}}{\lambda_{w} + \lambda_{R}}$ in the $\mathbf{k}\rightarrow{0}$ limit, we can use the DFT-spin texture and the known value of $\alpha$ to obtain $\lambda_{R}$ $\simeq$ -0.097 eV$\cdot$Angstrom and $\lambda_{w}$ $\simeq$ 0.1095 eV$\cdot$Angstrom, also reported in the main text.

Finally, the shape of the P3 Hamiltonian suggests the presence of a Weyl type of spin-crossing for the $\Bar{\Gamma}_{5}\oplus\Bar{\Gamma}_{6}$ states, which is absent in the high symmetry phase. This is confirmed by our calculation of the chirality (Chern number) at $\Gamma$ with the help of eq.~\ref{FE_k_dot_p} and the Z2pack code~\cite{PhysRevB.95.075146,PhysRevB.83.235401}. In particular, the evolution of the Wannier centre around a circular path centred at the $\Gamma$ point is traced for both the reference band and its time-reversed partner, revealing a non-trivial Z2 number.

\begin{figure}[ht]
     \includegraphics[width=0.7\textwidth]{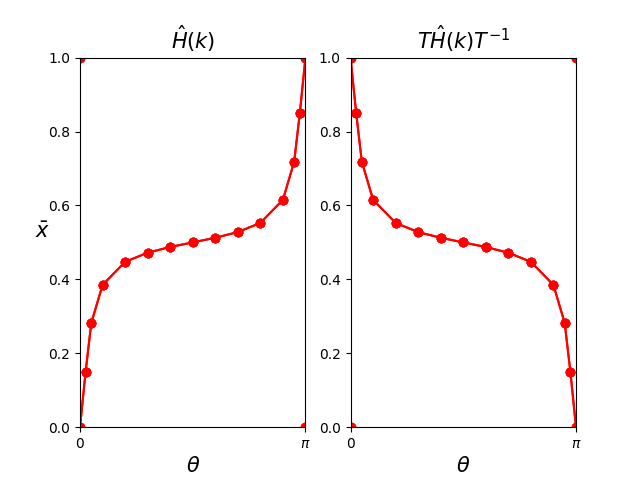}
     \caption{P3 phase $\Bar{\Gamma}_{5}\oplus\Bar{\Gamma}_{6}$ states, evolution of the Wannier centre $\bar{x}$ around a spherical loop centred at the $\Gamma$ point. The time reversal symmetry $T$ $\equiv$ i$\sigma_{y}$K (where K is the complex conjugation) accounts for the spin.}
     \label{Fig:WC_FE_band}
     \end{figure}

\subsection{PE-FE barrier, low-CB spin and band split-off terms with SOC off at selected Wyckoff positions}

When the SOC is deactivated on all Ge and O atoms, $\Delta$E is unchanged with respect to the full SOC case. This is consistent with the previous observations concerning the role of lead.
We observe that switching off the spin-orbit for the Pb-5d states produces no changes on the double well energy with respect to the full SOC-on case ($\sim$ 89 meV). On the other hand, deactivating the spin-orbit for the Pb-6p orbitals cancels the renormalisation of the energy well and brings $\Delta$E close to the no-SOC value ($\sim$ 68 meV). This confirms the role of the Pb-6p orbitals in this process.

\begin{figure}
 \centering
 \includegraphics[width=12cm,keepaspectratio=true]{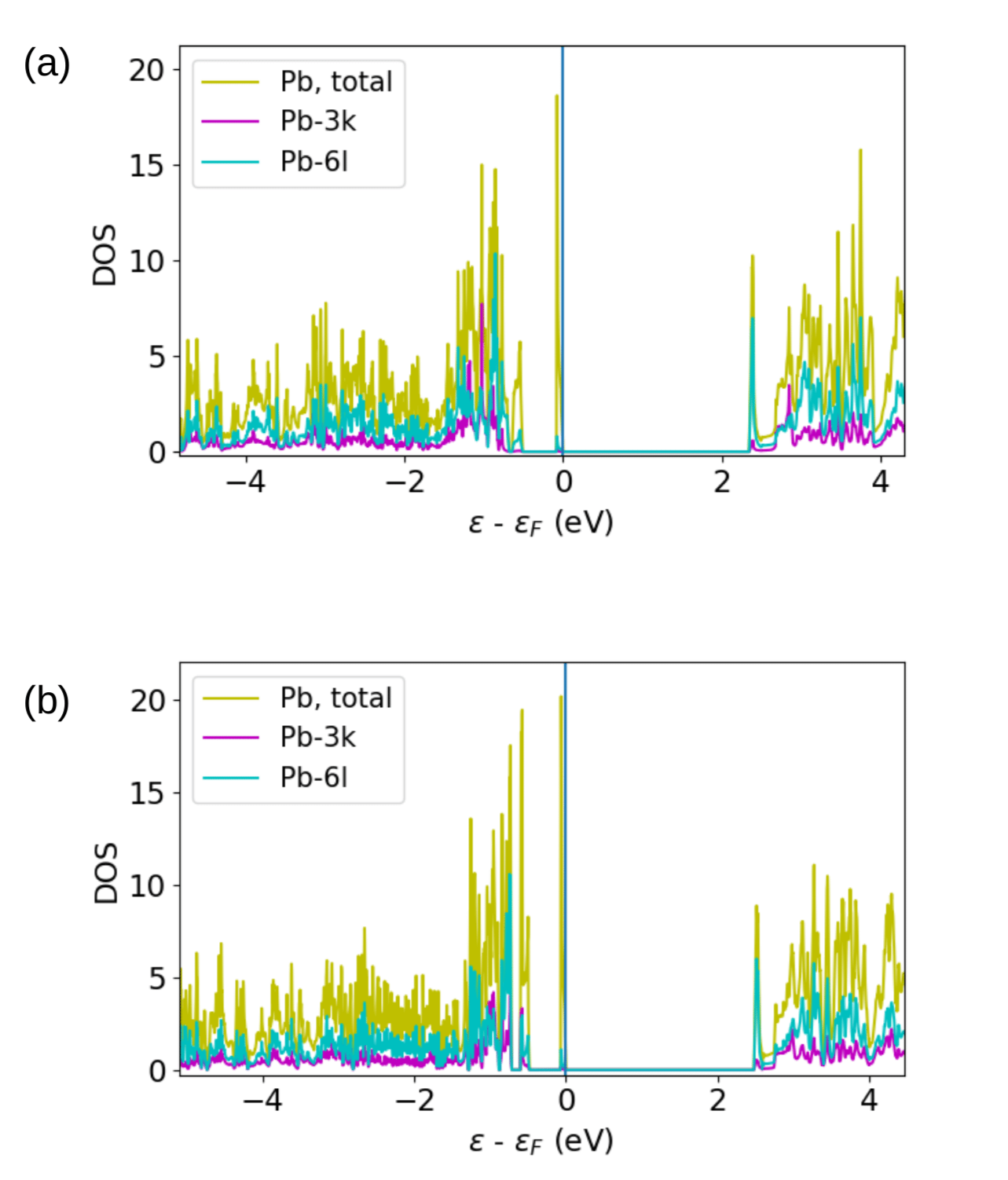}
\caption{Lead DOS contribution from the 3k and 6l positions. Both the P$\Bar{6}$ (a) and P3 (b) cases are shown.}
 \label{fig:DOS_Pb_3k_6l}
\end{figure}

Thus we can understand how much each lead atom contributes to the spin-orbital effects by deactivating the SOC at selected Wyckoff positions (WPs). 
We find that although all Pb atoms contribute to the shift, the sites with higher degeneracy show the largest effect. The $3k$ position is the one which, individually, contributes the most to the PE-FE energy difference, as we notice a negligible effect coming from the $2i$, $1c$ and $2h$ positions, since these positions are weakly degenerate. Moreover, we realise the presence of a competition between the spin-orbit at $3k$ and at $6l$ sites, since its deactivation in the latter case increases $\Delta$E. 
As stated in the main text, the $3k$-WP generates the mirror symmetry in the PE phase, which is broken by the transition. Thus it can be expected that such WPs provide a large contribution to the P$\Bar{6}$ $\rightarrow$ P3 energy landscape and with a strong SOC renormalisation. 
Since the other Pb-WPs possess unbroken site symmetries, the effect of the spin-orbit on the domain barrier is secondary and mainly due to covalency. We report the $3k$ and $6l$ spectral weights in fig.~\ref{fig:DOS_Pb_3k_6l}, which shows that both positions constitute much of the lead-PDOS and are likely to interact with neighbouring O atoms.


Moving on to P3-only quantities, we observe that deactivating the spin-orbit for $3k$ and $6l$ WPs strongly lowers the value of the E$_{R}$ and $\alpha$. The same effect is produced by switching off the SOC for all the other positions instead. However, we realise how the individual deactivation at $1e$, $1c$, $2h$ and $2i$ 
lead sites produces in fact a substantial increase of both $\alpha$ and E$_{R}$. For example, if the SOC is switched off at 2i sites, we find E$_{R}$ = 18.24 meV and $\alpha$ = 0.29 eV$\cdot$Angstrom, which is approximately twice the pristine value. 
The $\lambda_{\text{W}_{z}}$ is instead consistently small across all scenarios, as it is always found to be $\sim\leq$ 0.10 eV$\cdot$Angstrom.

\begin{widetext}
\begin{center}
\begin{table}[htbp!]
\begin{tabular}{cccccccccccc}
\hline
\hline
\multicolumn{8}{c}{SOC on:} &  $\Delta$E &  E$_{R}$ & $\alpha$ & $\lambda_{\text{W}_{z}}$  \\
Pb$_1$ (3k) & Pb$_2$ (6l) & Pb$_3$ (1e) & Pb$_4$ (2i) & Pb$_5$ (1c) & Pb$_6$ (2h) & Ge (all) & O (all) & \\
\hline
$\varnothing$ & $\varnothing$ & $\varnothing$ & $\varnothing$ & $\varnothing$ & $\varnothing$ & $\varnothing$ & $\varnothing$ &  68  & 0.00 & 0.00 & 0.00 \\
\checkmark & \checkmark & \checkmark & \checkmark & \checkmark & \checkmark & \checkmark & \checkmark &  89 & 3.40 & 0.15  &  0.01  \\
$\varnothing$ & $\varnothing$ & $\varnothing$ & $\varnothing$ & $\varnothing$ & $\varnothing$ & \checkmark & \checkmark &  67 &  0.00 &  0.00  &  0.00\\
\checkmark & \checkmark & \checkmark & \checkmark & \checkmark & \checkmark & $\varnothing$ & $\varnothing$ &  89 &  3.40 &  0.15 &  0.01 \\
$\varnothing$ & \checkmark & \checkmark & \checkmark & \checkmark & \checkmark & \checkmark & \checkmark &  76  &  0.69 &  0.06  & 0.09 \\
\checkmark & $\varnothing$ & \checkmark & \checkmark & \checkmark & \checkmark & \checkmark & \checkmark &  95 & 0.69 &  0.12 &  0.03 \\
\checkmark & \checkmark & $\varnothing$ & \checkmark & \checkmark & \checkmark & \checkmark & \checkmark &  83 & 21.45  & 0.27 &  0.06  \\
\checkmark & \checkmark & \checkmark & $\varnothing$ & \checkmark & \checkmark & \checkmark & \checkmark &  87 & 18.24 & 0.29 & 0.03 \\
\checkmark & \checkmark & \checkmark & \checkmark & $\varnothing$ & \checkmark & \checkmark & \checkmark &  89 & 9.50 & 0.20 & 0.02 \\
\checkmark & \checkmark & \checkmark & \checkmark & \checkmark & $\varnothing$ & \checkmark & \checkmark &  \color{black}85\color{black} & 5.93 & 0.10 & 0.03   \\
$\varnothing$ & $\varnothing$ & \checkmark & \checkmark & \checkmark & \checkmark & \checkmark & \checkmark &  \color{black}81\color{black} & 0.30 & 0.05 & 0.10 \\
\checkmark & \checkmark & $\varnothing$ & $\varnothing$ & $\varnothing$ & $\varnothing$ & \checkmark & \checkmark &  \color{black}76\color{black} & 0.13 & 0.04 & 0.11 \\
\hline
\end{tabular}
\label{tab:DeltaE-selected-SOC}
\caption{Gain of energy $\Delta$E (meV/f.u.) between the PE and the FE phases along with the E$_{R}$ (meV), $\alpha$ (eV$\cdot$Angstrom) and $\lambda_{\text{W}_{z}}$ (eV$\cdot$Angstrom) SOC-induced parameters for the lowest energy CB states ($\Bar{\Gamma}_{5}\oplus\Bar{\Gamma}_{6}$ representation, P3 phase). The $\checkmark$ ($\varnothing$) symbol means that SOC is (not) included for the considered P$\Bar{6}$ Wyckoff position. }
\end{table}
\end{center}
\end{widetext}

We have also calculated the band gap and the lowest two conduction bands splitting at $\Gamma$ generated by the spin-orbit coupling when activated at some WPs (for lead) and deactivated at other. Here with reference to fig. 2 in the main text we define $\delta \equiv |\text{E}(\Bar{\Gamma}_{5}\oplus\Bar{\Gamma_{6}}) - \text{E}(\Bar{\Gamma_{4}})|$ and $\gamma \equiv |\text{E}(\text{D}_{1/2}) - \text{E}(\Bar{\Gamma_{4}})|$ as descriptors. Naturally, in absence of SOC we have $\delta$ = 0 and $\gamma$ is bound from above by the $\Gamma_{1}-\Gamma_{3,5}$ crystal field splitting.
The results are shown in tab.~\ref{tab:SOC_WP_selected}.
We notice that activating the spin orbit interactions lowers the band gap, since it brings the $\Bar{E}$ (triplet) states closer to the Fermi level. Therefore, the bigger $\delta$, the bigger the band gap reduction.
At the same time, we realise that $\gamma$ tends to be negatively correlated with $|E_\text{gap}(\text{SOC})-E_\text{gap}(\text{no SOC})|$, since an increase of $\delta$ also tends to bring the $\Bar{\Gamma}_{4}$ and the J-singlet states close. Thus, a negative correlation is also encountered between $\delta$ and $\gamma$,  which we explain in terms of the weakening of Pb-O hybridisation (important for the phase transition) as the $\Bar{\Gamma}_{4}$ states are pushed higher in energy.

The positive correlation between $|E_\text{gap}(\text{SOC})-E_\text{gap}(\text{no SOC})|$ and the energy barrier between P3 domains $\Delta$E well agrees with the idea that the effect of the spin-orbit is transmitted to the valence states via Pb-6p/O-2p hybridisation.
Finally, we notice that $\delta$ and $\alpha$ tend to be negatively correlated.
A possible explanation of this effect could be the weakening of the Pb-O hybridisation upon increase of the CB-VB separation. As the $\Bar{\Gamma}_{4}$ states are pushed higher in energy, it is reasonable to expect that the interaction between O-2p and Pb-6p orbitals is reduced. This interaction likely favours the phase transition in PGO, so that its reduction has the side effect of making $\alpha$ smaller.
\begin{center}
\begin{table}[htbp!]
\begin{tabular}{ccccccccccc}
\hline
\hline
\multicolumn{8}{c}{SOC on:} &  E$_{\text{gap}}$ &  $\delta$ & $\gamma$ \\
Pb$_1$ (3k) & Pb$_2$ (6l) & Pb$_3$ (1e) & Pb$_4$ (2i) & Pb$_5$ (1c) & Pb$_6$ (2h) & Ge (all) & O (all) & \\
\hline
$\varnothing$ & $\varnothing$ & $\varnothing$ & $\varnothing$ & $\varnothing$ & $\varnothing$ & $\varnothing$ & $\varnothing$ &  2.48  & 0.0 & 270 \\
\checkmark & \checkmark & \checkmark & \checkmark & \checkmark & \checkmark & \checkmark & \checkmark &  2.25 & 180 & 106  \\ 
$\varnothing$ & $\varnothing$ & $\varnothing$ & $\varnothing$ & $\varnothing$ & $\varnothing$ & \checkmark & \checkmark &  2.48 & 0.0 &  270  \\ 
\checkmark & \checkmark & \checkmark & \checkmark & \checkmark & \checkmark & $\varnothing$ & $\varnothing$ &  2.25 &  180 &  106 \\
$\varnothing$ & \checkmark & \checkmark & \checkmark & \checkmark & \checkmark & \checkmark & \checkmark &  2.22  &  231 &  65 \\
\checkmark & $\varnothing$ & \checkmark & \checkmark & \checkmark & \checkmark & \checkmark & \checkmark &  2.19 & 343 &  135 \\
\checkmark & \checkmark & $\varnothing$ & \checkmark & \checkmark & \checkmark & \checkmark & \checkmark &  2.33 & 54  & 150 \\
\checkmark & \checkmark & \checkmark & $\varnothing$ & \checkmark & \checkmark & \checkmark & \checkmark &  2.32 & 53 & 186 \\
\checkmark & \checkmark & \checkmark & \checkmark & $\varnothing$ & \checkmark & \checkmark & \checkmark &  2.28 & 143 & 134 \\
\checkmark & \checkmark & \checkmark & \checkmark & \checkmark & $\varnothing$ & \checkmark & \checkmark &  \color{black}2.27\color{black} & 119 & 155\\ 
$\varnothing$ & $\varnothing$ & \checkmark & \checkmark & \checkmark & \checkmark & \checkmark & \checkmark &  \color{black}2.17\color{black} & 459 & 56 \\ 
\checkmark & \checkmark & $\varnothing$ & $\varnothing$ & $\varnothing$ & $\varnothing$ & \checkmark & \checkmark &  \color{black}2.30\color{black} & 250 & 57 \\
\hline
\end{tabular}
\label{tab:SOC_WP_selected}
\caption{Electronic band gap (eV), $\delta \equiv |\text{E}(\Bar{\Gamma}_{5}\oplus\Bar{\Gamma_{6}}) - \text{E}(\Bar{\Gamma_{4}})|$ and $\gamma \equiv |\text{E}(\text{D}_{1/2}) - \text{E}(\Bar{\Gamma_{4}})|$ (low-CB) splitting (meV). These quantities refer to the P3 phase only. The $\checkmark$ ($\varnothing$) symbol means that SOC is (not) included for the considered P$\Bar{6}$ Wyckoff position. }
\end{table}
\end{center}

\subsection{Electron doping and magnetism considerations}

\begin{figure}
    \centering
    \includegraphics[width = 11cm]{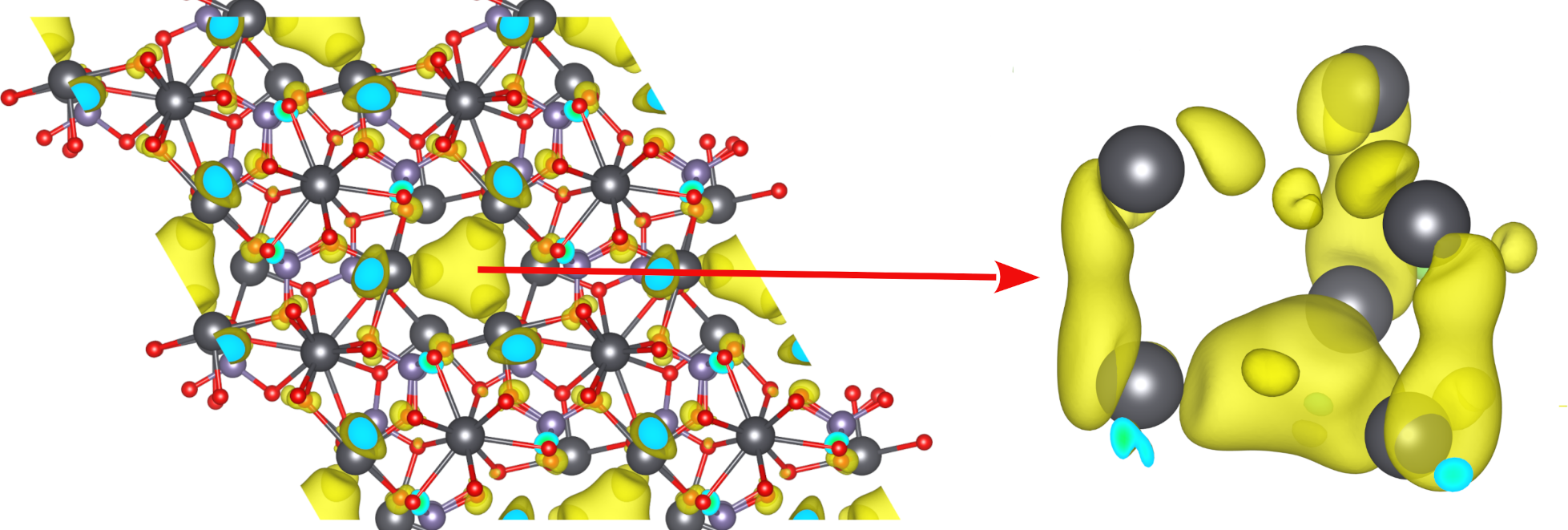}
    \caption{Calculated spin density (shown in yellow) with one extra electron in the ferroelectric phase. (left) Top view and (right) zoom on the cavity area with the 6l Pb atoms where 80\% of the additional electron state is present.}
    \label{fig:spin-density}
\end{figure}

To model negative carrier doping in PGO we change the total charge of the unit cell (ABINIT cellcharge flag).
Since the pressure of a charged system is ill-defined when periodic boundary conditions are employed~\cite{bruneval2015}, all the calculations are performed at fixed cell parameters, i.e. only the internal degrees of freedom are relaxed (atomic positions).
Considering a doping concentration of one extra electron per unit cell (FE phase), we compare the energy of the non-magnetic structure with that of a configuration where the magnetic moment is constrained to be nonzero. Also, we consider the no-SOC case for simplicity.
We find that the ferromagnetic solution is lower in energy by about 20 meV/f.u. with respect to the non-magnetic one. We also obtain that about 80\% of the electron is located inside the cavity state as seen in the spin-density fig.~\ref{fig:spin-density}. 
The depth of the ferroelectric double well is found to be of 87 meV/f.u. with one electron (in both the PE and the FE phases), i.e. 19 meV/f.u. more than without the extra electron.

\newpage

\end{document}